\documentclass[10pt, twocolumn]{IEEEtran}
\usepackage{stmaryrd}
\usepackage{stmaryrd}
\usepackage{bbm}
\usepackage{bbm}
\usepackage{mathrsfs}
\usepackage{amsfonts}
\usepackage{}
\makeatletter
\def\ps@headings{%
\def\@oddhead{\mbox{}\scriptsize\rightmark \hfil \thepage}%
\def\@evenhead{\scriptsize\thepage \hfil \leftmark\mbox{}}%
\def\@oddfoot{}%
\def\@evenfoot{}}
\makeatother
\usepackage{setspace}
\usepackage{amsfonts}
\usepackage{amssymb}
\usepackage{stmaryrd}
\usepackage{bbding}

\usepackage{amssymb}
\usepackage{algorithm}

\usepackage{paralist}
\usepackage{mdwlist}
\usepackage{algpseudocode}
\usepackage{multicol}
\usepackage{lipsum}
\usepackage{subfigure}
\usepackage{booktabs}
\usepackage{verbatim}
\usepackage{graphicx}
\usepackage{epstopdf}

\usepackage{amsthm}
\usepackage{multicol}

\theoremstyle{plain}
\newtheorem{mydef}{Definition}
\newtheorem{mytheorem}{Theorem}

\newcommand\Algphase[1]{%
\vspace*{-.2\baselineskip}\Statex\hspace*{\dimexpr-\algorithmicindent-0.5pt\relax}\rule{0.475\textwidth}{0.3pt}%
\Statex\hspace*{-\algorithmicindent}\textbf{#1}%
\vspace*{-.7\baselineskip}\Statex\hspace*{\dimexpr-\algorithmicindent-0.5pt\relax}\rule{0.475\textwidth}{0.3pt}%
}

\usepackage[cmex10]{amsmath}
\begin{document}

%
% paper title
% can use linebreaks \\ within to get better formatting as desired
\title{Making Double Spectrum Auction Practical:\\ Both Privacy and Efficiency Matter}

% author names and affiliations
% use a multiple column layout for up to three different
% affiliations

%%%%%%%%%%%%%%%%%%%%%%%%%%%%%%%%%% infocom 2015 requires not reveal author's names

%\author{\small \IEEEauthorblockN{Zhili Chen$^{1,2}$, Liusheng Huang$^{1,2}$, Lin Chen$^3$} \\
%\IEEEauthorblockA{$^1$School of Computer Science and Technology, University of Science and Technology of China, Hefei, China\\
%$^2$Suzhou Institute for Advanced Study, University of Science and Technology of China, Suzhou, China\\
%$^3$Lab. Recherche Informatique (LRI-CNRS UMR 8623), Univ. Paris-Sud, 91405 Orsay, France\\
%Email: \{zlchen3, lshuang\}@ustc.edu.cn, chen@lri.fr} }

\author{\IEEEauthorblockN{Zhili Chen, Xuemei Wei, Hong Zhong, Jie Cui, Yan Xu, Shun Zhang} \\
\IEEEauthorblockA{School of Computer Science and Technology, Anhui University, Hefei, China\\
Email: zlchen@ahu.edu.cn, shirlywei@outlook.com, \{zhongh,cuijie\}@mail.ustc.edu.cn,\\
xuyan@ahu.edu.cn, shzhang27@163.com} }

%\and \IEEEauthorblockN{Homer Simpson}
%\IEEEauthorblockA{Twentieth Century Fox\\
%Springfield, USA\\
%Email: homer@thesimpsons.com}
%\and
%\IEEEauthorblockN{James Kirk\\ and Montgomery Scott}
%\IEEEauthorblockA{Starfleet Academy\\
%San Francisco, California 96678-2391\\
%Telephone: (800) 555--1212\\
%Fax: (888) 555--1212}}

% conference papers do not typically use \thanks and this command
% is locked out in conference mode. If really needed, such as for
% the acknowledgment of grants, issue a \IEEEoverridecommandlockouts
% after \documentclass

% for over three affiliations, or if they all won't fit within the width
% of the page, use this alternative format:
%
%\author{\IEEEauthorblockN{Michael Shell\IEEEauthorrefmark{1},
%Homer Simpson\IEEEauthorrefmark{2},
%James Kirk\IEEEauthorrefmark{3},
%Montgomery Scott\IEEEauthorrefmark{3} and
%Eldon Tyrell\IEEEauthorrefmark{4}}
%\IEEEauthorblockA{\IEEEauthorrefmark{1}School of Electrical and Computer Engineering\\
%Georgia Institute of Technology,
%Atlanta, Georgia 30332--0250\\ Email: see http://www.michaelshell.org/contact.html}
%\IEEEauthorblockA{\IEEEauthorrefmark{2}Twentieth Century Fox, Springfield, USA\\
%Email: homer@thesimpsons.com}
%\IEEEauthorblockA{\IEEEauthorrefmark{3}Starfleet Academy, San Francisco, California 96678-2391\\
%Telephone: (800) 555--1212, Fax: (888) 555--1212}
%\IEEEauthorblockA{\IEEEauthorrefmark{4}Tyrell Inc., 123 Replicant Street, Los Angeles, California 90210--4321}}

% use for special paper notices
%\IEEEspecialpapernotice{(Invited Paper)}

%\author{\IEEEauthorblockN{Paper ID: 1570004869}}

% make the title area 1570004869
\maketitle

%\footnote{hello!}

\begin{abstract}
Truthful spectrum auction is believed to be an effective method for spectrum redistribution. However, privacy concerns have largely hampered the practical applications of truthful spectrum auctions. In this paper, to make the applications of double spectrum auctions practical, we present a privacy-preserving and socially efficient double spectrum auction design, SDSA. Specifically, by combining three security techniques: homomorphic encryption, secret sharing and garbled circuits, we design a secure two-party protocol computing a socially efficient double spectrum auction, TDSA, without leaking any information about sellers' requests or buyers' bids beyond the auction outcome. We give the formal security definition in our context, and theoretically prove the security that our design achieves. Experimental results show that our design is also efficient in performance, even for large-scale double spectrum auctions.

\end{abstract}
% IEEEtran.cls defaults to using nonbold math in the Abstract.
% This preserves the distinction between vectors and scalars. However,
% if the conference you are submitting to favors bold math in the abstract,
% then you can use LaTeX's standard command \boldmath at the very start
% of the abstract to achieve this. Many IEEE journals/conferences frown on
% math in the abstract anyway.

% no keywords

% For peer review papers, you can put extra information on the cover
% page as needed:
% \ifCLASSOPTIONpeerreview
% \begin{center} \bfseries EDICS Category: 3-BBND \end{center}
% \fi
%
% For peerreview papers, this IEEEtran command inserts a page break and
% creates the second title. It will be ignored for other modes.
\IEEEpeerreviewmaketitle

\section{Introduction}
% no \IEEEPARstart

%It is nowadays widely acknowledged that radio spectrum has become a
%precious resource, and the problem of spectrum scarcity is far more
%pressing than ever. Currently, the main causes of spectrum scarcity
%are two-fold: the increasing demand and the under-utilization. On
%one hand, more and more wireless devices and services are starving
%for new spectrum to use, making the available unlicensed spectrum a
%scarce resource. On the other hand, large bulks of licensed spectrum
%are inefficiently utilized by their current owners. To address the
%spectrum scarcity, a widely deployed solution is the spectrum
%redistribution which provides incentives to licensed users to trade
%their unused spectrum with unlicensed users. Recently, open markets
%for spectrum redistribution, such as Spectrum Bridge \cite{bridges},
%have begun to provide services for buying, selling, and leasing idle
%spectrum channels.

\subsection{Background and Motivation}

The dilemma between the rapid growth of wireless technologies and the scarcity of radio spectrum has given rise to the increasing demand for dynamic spectrum access. On the one hand, the spectrum scarcity is rooted in the limited spectrum resource, on the other hand it is aggravated by the traditional, static and rigid spectrum allocation policies by governments such as the Federal Communications Commission (FCC) in the United States. A survey~\cite{valenta2010survey} indicated that a significant amount of licensed spectrum remains unused in many places much of the time. Because of this, spectrum auctions allow primary users lease their idle channels for profit to secondary users who are starving for wireless spectrum, to improve the spectrum utilization, and thus alleviate the spectrum scarcity problem.

Recently, spectrum auctions have been studied extensively. Most of the existing works have focused on the achievement of truthfulness, stimulating each bidder to report its true valuation, under the implicit assumption that the auctioneer is trusted. However, if the auctioneer is actually dishonest, some security attacks may be launched by the auctioneer (or together with the corrupted bidders) to increase its profit while damaging the interests of the honest bidders \cite{11panm}. For example, the auctioneer can directly change the pricing to improve its revenue, or it can sell bidders' bid data to get profit. If such security issues are not appropriately addressed, the practical applications of spectrum auctions may be greatly hampered.

The source of the above-mentioned security attacks is that all bidders' bids are completely disclosed to the auctioneer in a spectrum auction without any security measure. To prevent these attacks, privacy preservation is called for, and in fact numerous works on privacy-preserving spectrum auctions have appeared. However, current works mainly focus on privacy-preserving single-sided spectrum auctions \cite{11panm}\cite{13huangq}\cite{15huangh}\cite{16huangq}\cite{chen2016towards}, while rarely address privacy-preserving double spectrum auctions \cite{14chenz}. The reason may be that privacy preservation in double spectrum auctions requires more complicated secure operations, and is more challenging than that in single-sided ones. Furthermore, though paper \cite{14chenz} indeed provided a secure double spectrum auction design PS-TRUST based on TRUST, from the view of practical applications, it still suffers the following two drawbacks. First, PS-TRUST incurred heavy overhead in term of computation and communication performance in large-scale auction scenarios. Second, PS-TRUST is based on TRUST, an early truthful double spectrum auction mechanism enabling spectrum reuse, but in practical applications, we may want to base on the improved version of TRUST, such as TDSA \cite{11yaoe}, while the direct transplant of PS-TRUST to TDSA is found to be insecure as later we will see. Therefore, a secure and efficient design for double spectrum auctions is still lacked.

\subsection{Contributions and Results}

\begin{figure}
  \centering
  % Requires \usepackage{graphicx}
  \includegraphics[width=0.47\textwidth]{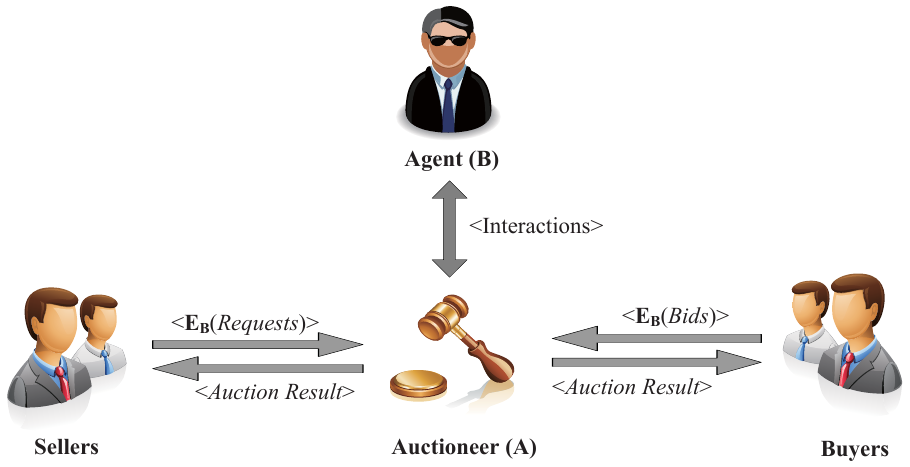}\\
  \caption{Auction Framework for SDSA}\label{fig:AuctionFramework}
\end{figure}

In this paper, we propose SDSA, Secure design for Double Spectrum Auctions based on TDSA, a more socially efficient double spectrum auction mechanism compared to TRUST. Our main idea is to introduce an agent who cooperates with the auctioneer to perform the secure auction computation, using the technology of secure two-party computation. In the auction framework as illustrated in Fig. \ref{fig:AuctionFramework}, each seller submits its request and each buyer submits its bid, encrypted by the agent's public key, to the auctioneer, who thus cannot decrypt the encrypted bids directly; then, with these encrypted requests and bids, the auctioneer and the agent perform the double spectrum auction cooperatively and securely, without any information about the requests or bids beyond the auction outcome leaked to anyone of them; finally, the auction outcome is released to the sellers and buyers by the auctioneer and the agent cooperatively. As long as the auctioneer and the agent do not collude, SDSA is proved to be secure.

Our main contributions can be described as follows.
\begin{itemize}
  \item We design a privacy-preserving yet practical double spectrum auction mechanism based on TDSA, combining the three security techniques, homomorphic encryption, secret sharing and garbled circuits, and formally prove that it is secure against semi-honest adversaries.
  \item To address the secure buyer group problem, we propose a novel technique, joint encryption, which encrypts a message between two parties, and discuss its nice properties for privacy preservation.
  \item To apply garbled circuits to the auction process so that both sellers' and buyers' privacies are well preserved, we design an elaborate and efficient data-oblivious algorithm of the underlying spectrum auction mechanism.
  \item We implement SDSA, and do extensive experiments to evaluate the performance in term of both computation and communication overheads.
\end{itemize}

\subsection{Related Work}\label{sec:relatedwork}

\textbf{Truthful double spectrum auctions.} Truthfulness attribute is usually emphasized in the design of double spectrum auctions. In the past few years, many mechanisms for truthful double spectrum auctions have been proposed. For example, Zhou et al. put forward TRUST \cite{09zhoux}, the first truthful double spectrum auction framework enabling spectrum reuse. Later, Yao et al. improved TRUST with an elaborated pricing mechanism, getting a more socially-efficient auction mechanism TDSA \cite{11yaoe}. Next, Chen et al. extended TRUST to enable multi-channel trades between sellers and buyers \cite{13chenz}. More recently, Chen et al. presented the first double auction design, TAMES \cite{14cheny}, for heterogeneous spectrum. Dong et al. proposed the first double spectrum auction design that explicitly decoupled the buyer-side and seller-side auction design while achieving economic-robustness \cite{16dongw}. However, all the studies mentioned above fail to consider the protection of bid privacy in the process of the auction.

\textbf{Secure traditional auctions.} In the past decade, a line of research has been devoted to secure or privacy-preserving auctions for traditional goods (e.g., stamps, antiques). The authors of \cite{99naorm} \cite{08brandtf} \cite{04yokoom} \cite{02pengk} employed various cryptographic techniques to design different secure auction mechanisms. Unfortunately, these secure traditional auction mechanisms failed to consider spectrum reuse, and thus they would lead to spectrum under-utilization when directly applied to spectrum allocation.

\textbf{Secure spectrum auctions.} Recently, more and more studies are devoted to secure spectrum auctions. Pan et al. \cite{11panm} provided the earliest spectrum auction design, starting this line of research. Later, the authors of \cite{13huangq} \cite{14chenz} \cite{14huangh} \cite{15chenz} \cite{16huangq} \cite{16suny} proposed various secure spectrum auction designs, with different emphasis. However, most of the existing studies focused on secure single-sided spectrum auctions, while secure double spectrum auctions were rarely addressed. Although both papers \cite{14chenz} and \cite{15chenz} indeed addressed secure double spectrum auction problems, paper \cite{14chenz} was distinguished from our work as pointed out in Section I-A, while paper \cite{15chenz} employed three-party computation, which requires a strong security assumption. In this paper, we will focus on the design of a secure, efficient and practical double spectrum auction mechanism, by leveraging on techniques of secure two-party computation.

\textbf{Our preliminary version.} A preliminary version [1] containing a subset of results of this paper has been presented in IEEE/ACM IWQoS 2017 held in Barcelona, Spain. Compared to the conference paper, this paper makes the following novel technical contributions:

\begin{enumerate}
\item \emph{Improved definition of joint encryption with randomization fields}: We improve the proposed technique, joint encryption, with randomization fields, describe its technical content in detail, and discuss its good properties for privacy preservation.
\item \emph{Detailed security proof of our design}: We prove the security of our design formally using the simulation based approach.
\item \emph{A brief review of related work}: We add a brief review of the related work, and point out the necessity of our work.
\item \emph{Design challenges and design rationale}: We add an illustration of design challenges(with an example), and a description of the design rationale to address these challenges.
\item \emph{Experimentation}: We have redone the experiments with our improved joint encryptions to evaluate its performance and demonstrate its practicality.
\end{enumerate}

Additionally, we have incorporated valuable feedback from the conference on our work, which results in significant improvement on both technical and editorial presentation.

\subsection{Paper Orgnization}

The remainder of this paper is organized as follows.
In Section~\ref{sec:problemstate}, the problem is formulated. Section
\ref{sec:preliminaries} provides technical preliminaries. We describe the design challenges and rationale in Section \ref{sec:sdsa-rationale}, and present the detailed design of SDSA and prove its security in Section
\ref{sec:sdsa-details}. In Section
\ref{sec:experiment}, we implement SDSA, and evaluate the performance in terms of computation and
communication overhead. Finally, the paper is concluded in Section
\ref{sec:conclusion}.

\section{Problem and Challenges}\label{sec:problemstate}

\subsection{Double Spectrum Auction Problem}

Double spectrum auction deals with the problem of how to organize trade between a set of secondary spectrum users (buyers) and a set of primary spectrum users (sellers). For example, double spectrum auction may be used as an effective approach by the FCC, which aims at reallocating spectrum licenses from TV broadcasters to mobile communication providers.

In our context, double spectrum auction is modeled as a sealed bid auction, where $M$ sellers want to sell their idle spectrum channels to $N$ buyers that need more spectrum. Notice that a spectrum auction is essentially different from a conventional auction in that spectrum can be reused by multiple buyers on the condition that they do not interfere with each other, and there are always far more buyers than sellers. In our double spectrum auction, each \emph{seller} $m$ submits its \emph{request} $Q_m$, while each \emph{buyer} $n$ submits its geographic location $(x_{n}, y_{n})$ and \emph{bid} $B_n$ to the \emph{auctioneer}, who evaluates the requests and bids, and determines the winning sellers and buyers and their respective clearing prices. For clarity, we assume that a seller provides one channel for sale and a buyer purchases one channel. All channels sold by sellers are homogeneous.

\subsection{TDSA Auction Mechanism}

To address the above-mentioned double spectrum auction problem, TRUST ~\cite{09zhoux} provided a representative truthful auction mechanism enabling spectrum reuse. Later, TDSA~\cite{11yaoe} improved TRUST to exploit better spectrum reusability, with all economic robustness well preserved. Specifically, TDSA consists of the following five steps.

(1) \textbf{Buyer group formation}: The auctioneer privately constructs a conflict graph of buyers according to their geographic locations, and forms non-conflict and bid-independent buyer groups subject to the conflict graph. The buyer groups are denoted as $G_{t}$, $t\in\{1,2,\ldots,T\}$.

(2) \textbf{Virtual group bidding}: Assume there are $R_{t}$ members in group $G_{t}$. For each group $G_{t}$, TDSA first sorts bids of buyers in non-increasing order: $B_{t,1} \geq B_{t,2} \geq \cdots \geq B_{t,R_{t}}$. Let the set $\textit{VG}_{j} = \{ B_{t,1}, B_{t,2},\ldots, B_{t,j} \}$ be a virtual group whose virtual group bid is $ B_{t,j}^{v} = B_{t,j} \times j $. Then, group $G_{t}$ has $R_{t}$ virtual groups denoted as $\textit{VG}_{j}$, $j\in\{ 1,2,\ldots,R_{t}\}$. The bid of group $G_{t}$ could be described as $B_{t}^{g}=\max_{j}(B_{t,j}^{v})$.
In the next step, each buyer group $G_{t}$ will act as a single ``buyer'' bidding its group bid $B_{t}^{g}$.

(3) \textbf{Preliminary winner determination}: McAfee's design is applied in this step. First, sort the requests of sellers in non-decreasing order, and sort the group bids of buyer groups in non-increasing order.

\begin{center}
$O':Q_{1}\leq Q_{2} \leq \ldots \leq Q_{M}$
\end{center}
\begin{center}
$O'':B_{1}^{g}\geq B_{2}^{g} \geq \ldots \geq B_{T}^{g}$
\end{center}

Then, find the last profitable index as follows:
%$k=\arg\max_{l\leq\min\{T,M\}}\{GBid_{l} \geq Q_{l}$ and $GBid_{l} \neq GBid_{l-1}\}$
%\begin{displaymath}
%  k = \textrm{arg max}_{l\leq\min\{T,M\}} \{GBid_{l} \geq Q_{l} \, and \, GBid_{l}\neq GBid_{l-1}\}
%\end{displaymath}
\begin{equation*}
   \Phi = {\arg\max}_{\phi\leq\min\{T,M\}} \{B_{\phi}^{g} \geq Q_{\phi} \, and \, B_{\phi}^{g}\neq B_{\phi-1}^{g}\}
\end{equation*}

The winning sellers are the top $\Phi-1$ sellers in $O'$, and the clearing price for them is the $\Phi_{th}$ ranked seller's request ($P^{s}=Q_{\Phi}$). Similarly, the preliminary winning buyer groups are the top $\Phi-1$ buyer groups in $O''$, and the corresponding clearing price is the $\Phi_{th}$ ranked buyer group's bid ($P^{g}=B_{\Phi}^{g}$).

(4) \textbf{Washing out}: This step is carried out in each winning buyer group to remove buyers whose bids are too low to afford the clearing price. For each winning group $G_{t}$, TDSA first sorts buyers' bids in non-increasing order, $B_{t,1} \geq B_{t,2} \geq \cdots \geq B_{t,R_{t}}$, and then washes out buyers in $G_t$ one by one from $R_{t}$ to $1$, until it finds a virtual group $\textit{VG}_{C_t} = \{ B_{t,1}, B_{t,2},\ldots, B_{t,C_t} \}$ whose virtual bid is higher than the clearing price for winning buyer groups (i.e., $B_{t,C_{t}}^{v}=B_{t,C_{t}} \times C_{t}>P^{g}$).

(5) \textbf{Final pricing}: The auctioneer buys a channel from each winning seller at the price of $P^{s}$, and sells a distinct channel to each winning group at the price of $P^g$. Furthermore, in a winning group $G_t$, each winning buyer equally pays $\frac{P^g}{C_t}$.

In the auction, the symbols used are summarized in Tab.~\ref{tab:KeySymbolsforSDSA}.

\begin{table}[htbp]\small
  \caption{\small Key Symbols for SDSA}
  \label{tab:KeySymbolsforSDSA}
  \centering
  \begin{tabular}{rl}
    \hline
    % after \\: \hline or \cline{col1-col2} \cline{col3-col4} ...
    $M$, $N$ & numbers of sellers and buyers \\
    $T$ & numbers of buyer groups \\
    $Q_{m}$, $B_{n}$ & request of seller $m$, and bid of buyer $n$ \\
    $(x_{n}, y_{n})$ & location of buyer $n$ \\
    $B_{t}^{g}$ & group bid of group $t$ \\
    $R_{t}$ & numbers of buyers within group $t$ \\
    $C_{t}$ & numbers of winning buyers within group $t$ \\
    $S$ & $S=\{Q_{i}\}_{i=1}^{M}$, global bid set of sellers \\
    $G_{t}$ & $G_{t}=\{B_{t}^{g},\{B_{t,j}\}_{j=1}^{R_{t}}\}$, per-group bid set \\
    $G$ & $G=\{G_{t}\}_{t=1}^{T}$, global bid set of groups \\
    $P^{s}$ & clearing price for winning sellers \\
    $P^{g}$ & clearing price for winning buyer groups \\
    $P^{b}_{n}$ & clearing price for winning buyer $n$ \\
    \hline
  \end{tabular}
\end{table}

\subsection{Security Requirements}\label{sec:security-req}

Our aim is to design a privacy-preserving double spectrum auction mechanism (SDSA) based on TDSA in the semi-honest adversary model, such that this mechanism reveals to any party nothing about the sensitive inputs (e.g., all sellers' requests, and all buyers' bids) except the auction outcome. Here, by any party, we mean:
\begin{itemize}
  \item The auctioneer;
  \item The agent, who is introduced for secure computations;
  \item The sellers and buyers;
  \item And anyone not participating in the auction.
\end{itemize}
By the auction outcome, we mean:
\begin{itemize}
  \item All winning sellers and their clearing prices;
  \item All winning buyer groups and their clearing prices.
\end{itemize}

Ideally, our secure mechanism requires that, through the auction, any party can only learn the auction outcome, but nothing else about the sensitive inputs. This can be simply verified for the sellers and buyers, and anyone not participating in the auction, since they do not take part in the auction computations, and what they get from the auction is the published auction outcome. However, it is much more complicated to prove this for the auctioneer or the agent, because all secure computations are performed between them. We define formally the computation security in our context, in term of two-party computation security definition \cite{04goldreicho}, as follows.

\begin{mydef}[\textbf{Security against semi-honest adversaries}]
Let $f(x,y)$ be a two-party auction functionality with the inputs $x$ and $y$ from the two parties, respectively, and an auction
outcome $f(x,y)$. Suppose that protocol $\Pi$
computes functionality $f(x,y)$ between the auctioneer and the agent.
Let $V^{\Pi}_A(x,y)$ (resp. $V^{\Pi}_B(x,y)$) represent the auctioneer's
(resp. the agent's) view during an execution of $\Pi$ on $(x,y)$. In other
words, if $(x, \textbf{r}^{\Pi}_A)$ (resp. $(y,\textbf{r}^{\Pi}_B)$)
denotes the auctioneer's (resp. the agent's) input and randomness, then
\[
\begin{array}{l}
V^{\Pi}_A(x,y) = (x, \textbf{r}^{\Pi}_A, m_1, m_2,...,m_t),\text{ and} \\
V^{\Pi}_B(x,y) = (y, \textbf{r}^{\Pi}_B, m_1, m_2,...,m_t)
\end{array}
\]
where $\{m_i\}_{i=1}^t$ denote the messages passed between the two parties. Let
$O^{\Pi}(x,y)$ denote the auction outcome after an execution of $\Pi$ on $(x,y)$. Then we say that protocol $\Pi$
is \textbf{secure} (or \textbf{preserves privacy}) against semi-honest adversaries if
there exist probabilistic polynomial time (PPT) simulators $S_1$ and
$S_2$ such that
\begin{equation}\label{equ:semisecure1}
\{(S_1(x, f(x,y)), f(x,y))\} \overset{c}{\equiv}
\{(V^{\Pi}_A(x,y), O^{\Pi}(x,y))\}
\end{equation}
\begin{equation}\label{equ:semisecure2}
\{(S_2(y, f(x,y)), f(x,y))\} \overset{c}{\equiv}
\{(V^{\Pi}_B(x,y), O^{\Pi}(x,y))\}
\end{equation}
where $\overset{c}{\equiv}$ denotes computational
indistinguishability.
\end{mydef}

Thus, the security requirements are essentially that the auction computation between the auctioneer and the agent should satisfy the above security definition.

\section{Technical Preliminaries}
\label{sec:preliminaries}

In this section, we introduce technical preliminaries, i.e., garbled circuits, Pallier cryptosystem and joint encryption, for our design.

\subsection{Garbled Circuit}
\label{sec:preliminaries:garbledcircuit}
Garbled circuits~\cite{Yao1982ProtocolsFS}~\cite{Lindell2009APO} (a.k.a. Yao's protocol) is an efficient method for general secure two-party computation. Suppose that two parties $A$ and $B$
want to compute a joint functionality $f(x,y)$ (represented by a Boolean circuit) with $A$'s input $x$ and $B$'s input $y$. The basic idea of garbled circuits is that one party, say $A$, constructs a garbled circuit from $f(x,y)$, and the other party, say $B$, computes the garbled circuit without learning any input and intermediate values. Garbled circuits protect privacy so long as the two parties do not collude with each other. The protocol proceeds as follows.

\textbf{Garbled circuit construction.} Party $A$ constructs a garbled circuit consisting of multiple binary gates, each of which has two input wires and one output wire, as below. First, $A$ chooses two random cryptographic keys, $K_{w_{i}}^{0}$ and $K_{w_{i}}^{1}$, for each wire $w_{i}$ of the circuit ($K_{w_{i}}^{0}$ encodes a 0-bit, and $K_{w_{i}}^{1}$ encodes a 1-bit, but $K_{w_{i}}^{\sigma}$ does not reveal $\sigma$). Thus, for each gate g (e.g., an AND gate as illustrated in Tab. \ref{tab:and}), six keys $K_{w_{i}}^{0}$,$K_{w_{i}}^{1}$,$K_{w_{j}}^{0}$,$K_{w_{j}}^{1}$,$K_{w_{k}}^{0}$,$K_{w_{k}}^{1}$ will be obtained corresponding to the two input wires $w_{i}$, $w_{j}$ and the output wire $w_{k}$. Next, $A$ computes the four ciphertexts
\begin{center}
$\text{Enc}_{K_{w_{i}}^{b_{i}},K_{w_{j}}^{b_{j}}}(K_{w_{k}}^{g(b_{i},b_{j})})$ for $b_{i},b_{j}\in \{0,1\}$
\end{center}
These four ciphertexts are then randomly ordered, and constitute a garbled gate $g$. The collection of all garbled gates forms the garbled circuit, and a corresponding decoding table, which converts the output keys of the garbled circuit to real bits, is created. Finally, $A$ sends both the garbled circuit and the decoding table to $B$.
\begin{table}[htbp]\small
  \caption{A garbled AND gate: $g(b_i, b_j)=b_i \wedge b_j$}
  \label{tab:and}
  \centering
  \begin{tabular}{cccc}
    \hline
    $b_i \leftrightarrow w_i$ & $b_j \leftrightarrow w_j$ & $g(b_i, b_j) \leftrightarrow w_k$ & Garbled value \\
    \hline
    $0 \leftrightarrow K^0_{w_i}$ & $0 \leftrightarrow K^0_{w_j}$ & $0 \leftrightarrow K^0_{w_k}$ & $\text{Enc}_{(K^0_{w_i}, K^0_{w_j})}(K^0_{w_k})$ \\
    $0 \leftrightarrow K^0_{w_i}$ & $1 \leftrightarrow K^1_{w_j}$ & $0 \leftrightarrow K^0_{w_k}$ & $\text{Enc}_{(K^0_{w_i}, K^1_{w_j})}(K^0_{w_k})$ \\
    $1 \leftrightarrow K^1_{w_i}$ & $0 \leftrightarrow K^0_{w_j}$ & $0 \leftrightarrow K^0_{w_k}$ & $\text{Enc}_{(K^1_{w_i}, K^0_{w_j})}(K^0_{w_k})$ \\
    $1 \leftrightarrow K^1_{w_i}$ & $1 \leftrightarrow K^1_{w_j}$ & $1 \leftrightarrow K^1_{w_k}$ & $\text{Enc}_{(K^1_{w_i}, K^1_{w_j})}(K^1_{w_k})$ \\
    \hline
  \end{tabular}
\end{table}

%\begin{figure}[htbp]
%  \centering
%  % Requires \usepackage{graphicx}
%  \includegraphics[width=0.5\textwidth]{file/ANDgate.pdf}\\
%  \caption{A garbled AND gate}\label{fig:ANDgate}
%\end{figure}

\textbf{Input garbling.} Having constructed the garbled circuit and the decoding table, $A$ also has to garble both inputs $x$ and $y$. For input $x$, i.e., $A$'s input, $A$ can simply encode it into random keys $G(x)$, and send $G(x)$ to $B$. For input $y$, i.e., $B$'s input, $B$ cannot directly send $y$ to $A$ for garbling due to the privacy of $y$, but both $A$ and $B$ engage in a 1-out-of-2 oblivious transfer protocol, and thus $B$ also gets random keys $G(y)$ corresponding to $y$.

\textbf{Garbled circuit computation.} Upon receiving the garbled circuit, the decoding table, and the garbled inputs $G(x)$ and $G(y)$, $B$ computes the garbled circuit with $G(x)$ and $G(y)$, obtains the garbled result $G(f(x,y))$, and subsequently decodes it with the decoding table to get the output $f(x,y)$ in the clear. Finally, if necessary, $B$ will send the common result $f(x,y)$ to $A$.

\subsection{Paillier Cryptosystem}
\label{sec:preliminaries:pailliercryptosystem}

Paillier cryptosystem ($G$, $E$, $D$)~\cite{paillier1999public} is a widely-used homomorphic encryption, where $G$, $E$ and $D$ denote the key generation, encryption and decryption algorithms, respectively.

\noindent\textbf{Key generation:} $(pk,sk) = G(1^{\gamma })$.
\begin{enumerate}[1)]%\indent 1.
   \item Choose two large random prime numbers $p$ and $q$ and independently of each other such that $\text{gcd}(pq,(p-1)(q-1))=1$.
   \item Compute $n=pq$, $\lambda=$ lcm$(p-1,q-1)$.
   \item Select random integer $g$ where $g\in_{R} \mathbb{Z}_{n^{2}}^{*}$.
   \item Compute $\mu=(L(g^{\lambda}$ mod $n^{2}))^{-1}$ mod $n$ if exist, where $L(x)=(x-1)/n$.
   \item The public key $pk$ is $(n,g)$ and the private key is $(\lambda, \mu)$.
\end{enumerate}

\noindent\textbf{Encryption:}  $c=E_{pk}(m,r)$.
\begin{enumerate}[1)]%\indent 1.
   \item Let $m$ be a message to be encrypted where $m\in \mathbb{Z}_{n}$.
   \item Select random $r$ where $r\in \mathbb{Z}_{n}$.
   \item Compute ciphertext as $c= g^{m}\cdot r^{n}$ mod $n^{2}$.
\end{enumerate}

\noindent\textbf{Decryption:}  $m=D_{sk}(c)$.
\begin{enumerate}[1)]%\indent 1.
   \item Let $c$ be the ciphertext to decrypt where $c\in\mathbb{Z}_{n^{2}}^{*}$.
   \item Compute ciphertext as $m=L(c^{\lambda}$ mod $n^{2})\cdot \mu$ mod $n$.
\end{enumerate}

Given messages $m_{1}$, $m_{2}$, $m$, random numbers $r_{0}$, $r_{1}$, $r_{2}$ and $r$, and the integer $n$, which is the product of two large primes, i.e., $n = p \times q$, we can conclude the three technical properties of Paillier cryptosystem as follows.

(1) \emph{Indistinguishability}. Considering a message $m$ is encrypted twice, these two ciphertexts $E(m,r_{1})$ and $E(m,r_{2})$ are thoroughly independent, and no one can determine whether they correspond to the same message with a probability significantly higher than random guess without decrypting them.

(2) \emph{Homomorphic addition}. The product of two ciphertexts will decrypt to the sum of their corresponding messages, and the $k$th power of a ciphertext will decrypt to the product of the message and $k$, where $k$ is a constant.
%\[
%D(E(m_{1},r_{1}) \cdot E(m_{2},r_{2}) \!\mod n^{2}) = m_{1}+m_{2} \!\mod n
%\]
%\[
%D(E(m,r)^{k} \!\mod n^{2}) = k \cdot m \!\mod n
%\]

\begin{center}
$D(E(m_{1},r_{1}) \cdot E(m_{2},r_{2}) \!\mod n^{2}) = m_{1}+m_{2} \!\mod n$
\end{center}
\begin{center}
$D(E(m,r)^{k} \!\mod n^{2}) = k \cdot m \!\mod n$
\end{center}

%\begin{align*}
%  D(E(m_{1},r_{1}) \cdot E(m_{2},r_{2}) \, \textrm{ mod } \, n^{2}) & = m_{1}+m_{2} \, mod \, n \\
%  D(E(m,r)^{k} \, mod \, n^{2}) & = k \cdot m \, mod \, n
%\end{align*}

(3) \emph{Self-blinding}. Any ciphertext can be publicly changed into another one without affecting the message. Specifically, a different randomized ciphertext $E(m,r')$ can be computed from the ciphertext $E(m,r)$ by multiplying a ciphertext of zero, but without decryption and re-encryption.
%\begin{align*}
%   D(E(m,r)\cdot E(0,r_{0}) \mod n^{2}) & = D(E(m+0,r')) \\
%   & = D(E(m,r')) \\
%   & = m \mod n,
%\end{align*}
\begin{align*}
   D(E(m,r)\cdot E(0,r_{0}) \!\!\!\mod n^{2}) & = D(E(m+0,r')) \\
   & = D(E(m,r')) \\
   & = m \!\!\!\mod n,
\end{align*}
where $E(m,r')\neq E(m,r)$.

Notice that every encryption $E(m,r)$ of Paillier cryptosystem requires a random number $r$, which allows the randomized encryption. For simplicity, we will omit $r$ in the paper that follows, and denote $E(m,r)$ with public key $pk$ as $[m]_{pk}$, while bear in mind the randomness.

\subsection{Joint Encryption}\label{sec:preliminaries:jointencryption}

\begin{figure}
  \centering
  % Requires \usepackage{graphicx}
  \includegraphics[width=0.3\textwidth]{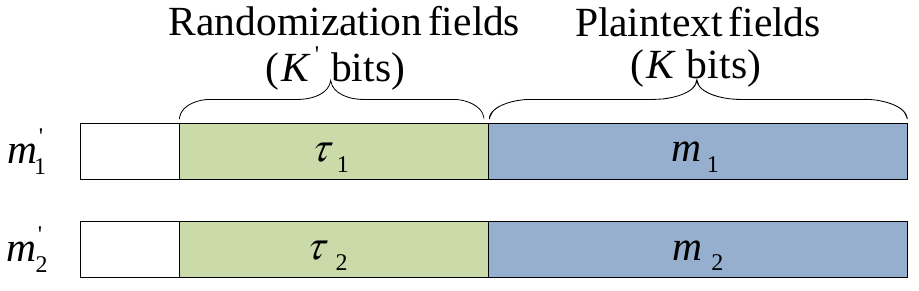}\\
  \caption{Joint encryption of a plaintext $m$}\label{fig:jencrypt}
\end{figure}

On the basis of the Paillier cryptosystem, we design the \emph{joint encryption}, which allows two parties to jointly encrypt a message without learning the message. Furthermore, joint encryption allows any party to randomize a joint ciphertext, such that any other parity (including the two encrypting parties) cannot link the randomized form to its original form. Let ($pk_{1}$, $sk_{1}$) and ($pk_{2}$, $sk_{2}$) denote the public/private key pairs of party $1$ and party $2$, respectively. $n_{1}$ and $n_{2}$ represent the corresponding products of two large primes. The joint encryption of a message $m$ is defined as (see Fig.~\ref{fig:jencrypt}):
\begin{equation}\label{equ:JointEncryption}
\begin{split}
    \langle (\![m_1]\!)_{pk_1}, (\![m_2]\!)_{pk_2} \rangle &= \langle [m_1^{\prime}]_{pk_{1}},[m_2^{\prime}]_{pk_{2}}\rangle \\
                                                           &= \langle [m_1 + \tau_1]_{pk_{1}},[m_2 + \tau_2]_{pk_{2}}\rangle
\end{split}
\end{equation}
where $m_1 + m_2 = m \pmod{2^{K}}$ with $m$, $m_{1}$, $m_{2} \in \mathbb{Z}_{2^{K}}$, and $K$ is the bit number used to represent the plaintext fields; $\tau_1, \tau_2 \in [2^K, 2^{K+K^{\prime}})$, and $K^{\prime}$ is the bit number for the randomization fields; obviously, $0<2^{K + K^{\prime}} < \min(n_{1},n_{2})$.

Particularly, a secret value $x\in\mathbb{Z}_{2^{K}}$ held by an external user can be jointly encrypted by the two parties as follows.
\begin{enumerate}
  \item The external user encrypts $x$ with the public key of a party, say party $2$, and sends the resulted ciphertext $[x]_{pk_{2}}$ to another party, party $1$.
  \item Party $1$ selects random number $s\in\mathbb{Z}_{2^{K}}$ uniformly, then divides $x$ into two shares by computing
      \begin{align}
        &[x_1]_{pk_1} \leftarrow [s]_{pk_{1}} \text{ mod } n_{1}^{2} \text{ and } \label{formula:x_1}\\
        &[x_2]_{pk_2} \leftarrow [x]_{pk_{2}}\cdot [2^{K}-s]_{pk_{2}} \text{ mod } n_{2}^{2} \label{formula:x_2}.
      \end{align}
  \item Party $1$ modifies $x_{1}$ and $x_{2}$ with randomization fields $\tau_{1}\in_{R}[2^K,2^{K + K^{\prime}})$ and $\tau_{2}\in_{R}[2^K,2^{K + K^{\prime}})$, respectively, where $K_{1}$ and $K_{2}$ represent the bit numbers of $n_{1}$ and $n_{2}$, respectively, and $K^{\prime}$ represents the bit number for randomization fields.
      \begin{align}
        &(\![x_1]\!)_{pk_1} \leftarrow [x_{1}]_{pk_{1}}\cdot [\tau_{1}]_{pk_{1}} \text{ mod } n_{1}^{2} \\
        &(\![x_2]\!)_{pk_2} \leftarrow [x_{2}]_{pk_{2}}\cdot [\tau_{2}]_{pk_{2}} \text{ mod } n_{2}^{2}
      \end{align}
      Finally, it gets the joint encryption $\langle(\![x_1]\!)_{pk_{1}}, (\![x_2]\!)_{pk_{2}}\rangle$.
\end{enumerate}

The randomization fields $\tau_{1}$ and $\tau_{2}$ above are indispensable. To understand the purpose of them, consider the case when the secret shares $x_{1}$ and $x_{2}$ in plaintext fields are reallocated, e.g., Eq.~\eqref{formula:x_1} and \eqref{formula:x_2} (Note that the shares are reallocated from $\langle 0, x \rangle$ to $\langle x_1, x_2 \rangle$). Specifically, in Eq.~\eqref{formula:x_2}, when the two K-bit numbers $x$ and $(2^K - s)$ are added together in the plaintext space, there may be a carry bit 1 to the $(K+1)$th bit. If nothing is done with this carry bit, then it can be used to link the resulted form of $x$ to its original form. For example, if there is a carry bit 1 in Eq.~\eqref{formula:x_2}, then we can link the resulted joint ciphertext to a larger $x$ with more probability. To hide this carry bit, we add random numbers $\tau_1, \tau_2 \in_R [2^K, 2^{K+K^{\prime}})$ to the randomization fields. Since the carry bit is 0 or 1, then the resulted distributions of randomization fields are $[2^K, 2^{K+K^{\prime}})$ and $[2^K + 1, 2^{K+K^{\prime}} + 1)$, which are indistinguishable from each other only with negligible exception (i.e., $2^{1-K^{\prime}}$) if $K^{\prime}$ is of the order of the security parameter.

If the two parties are clear in the context, we denote $[\![x]\!]=\langle(\![x_1]\!)_{pk_{1}}, (\![x_2]\!)_{pk_{2}}\rangle$ for short, and call $[\![x]\!]$ a \emph{joint ciphertext} of $x$. Note that $[\![x]\!]$ can be held by both party 1 and party 2, but none of them can learn any information about $x$ as long as they do not collude with each other. Furthermore, if each party $i$ decrypts its corresponding part $[\![x]\!]_i = (\![x_{i}]\!)_{pk_{i}}$, they get their respective additive shares $x_i$ of $x$.

The joint encryption based on Paillier cryptosystem satisfies the same properties (i.e., indistinguishability, homomorphic addition, and self-blinding) as Paillier cryptosystem for parties other than the encrypting parties, as well as an additional property (i.e., strong self-blinding) even for the encrypting parties. We describe the properties as follows.

(1) \emph{Indistinguishability}. This property holds straightforward due to the fact that a joint encryption is composed of two Paillier encryptions.

(2) \emph{Homomorphic addition}. Let $[\![x]\!]$ and $[\![y]\!]$ be the joint encryptions of messages $x$, $y\in \mathbb{Z}_{2^{K}}$ and let $k\in \mathbb{N}$ be a public constant. Joint ciphertext $[\![x+y]\!]$ and $[\![kx]\!]$ can be computed as follows.
\begin{align*}
   & D_{(pk_{1},pk_{2})}([\![x]\!] \cdot [\![y]\!]) \\
   & = \langle D_{pk_{1}}((\![x_{1} + y_{1}]\!)_{pk_{1}}), D_{pk_{2}}((\![x_{2} + y_{2}]\!)_{pk_{2}})\rangle \\
   & = (x + y) \pmod{2^{K}} \\
   & D_{(pk_{1},pk_{2})}([\![x]\!]^k) \\
   & = \langle D_{pk_{1}}((\![k \cdot x_{1}]\!)_{pk_{1}}), D_{pk_{2}}((\![k \cdot x_{2}]\!)_{pk_{2}})\rangle \\
   & = k \cdot x  \pmod{2^{K}}
\end{align*}

(3) \emph{Self-blinding}.
This property can be easily derived from the homomorphic addition property of Pallier cryptosystem.

(4) \emph{Strong self-blinding}.
Last but not least, we denote the important additional property of our joint encryption as \emph{strong self-blinding}, which allows any party 1 or 2 to change a joint ciphertext into another form. Different from self-blinding, strong self-blinding, \emph{SS-blinding} for short, changes not only the random parameters in two Paillier ciphertexts but also the secret shares with respect to each party. Specifically, SS-blinding can be carried out by party $i (i=1,2)$ as below.

For a joint ciphertext $[\![x]\!]$, party $i$ first computes $[\![0]\!] = \langle (\![\nu]\!)_{pk_{1}}, (\![2^K-\nu]\!)_{pk_{2}}\rangle$, where $\nu\in_R \mathbb{Z}_{2^{K}}$. Then, party $i$ changes the form of $[\![x]\!]$ by adding 0 to the plaintext using homomorphic addition as follows.
\begin{align*}
   & [\![x]\!] \cdot [\![0]\!] = \langle (\![x_{1}]\!)_{pk_{1}}, (\![x_{2}]\!)_{pk_{2}}\rangle \cdot \langle (\![\nu]\!)_{pk_{1}}, (\![2^K - \nu]\!)_{pk_{2}}\rangle\\
   & = \langle (\![x_{1} + \nu]\!)_{pk_{1}}, (\![x_{2} + 2^K - \nu]\!)_{pk_{2}}\rangle \\
   & = \langle (\![x^{(1)}_{1}]\!)_{pk_{1}}, (\![x^{(1)}_{2}]\!)_{pk_{2}}\rangle \\
   & = [\![x]\!]^{(1)}
\end{align*}
where $x^{(1)}_{1}+x^{(1)}_{2}=x \pmod{2^K}$ still holds. We denote the resulted form by $[\![x]\!]^{(1)}=\langle (\![x^{(1)}_{1}]\!)_{pk_{1}}, (\![x^{(1)}_{2}]\!)_{pk_{2}}\rangle$, which is another joint encryption form of the original message $x$. It is obvious that a joint ciphertext $[\![x]\!]$ can be SS-blinded many times if necessary, and we denote the $\varphi$th SS-blinding of $[\![x]\!]$ as $[\![x]\!]^{(\varphi)}$. Intuitively, when it comes to a specific protocol containing many joint ciphertexts, SS-blinding from party $i$ prevents the other party from linking a joint ciphertext with its SS-blinded forms.

Although we only focus on two parties, our joint encryption can be easily extended to multiple parties.

\section{SDSA: Design Challenges and Rationale}\label{sec:sdsa-rationale}
In this section, we describe the design challenges and rationale of SDSA.

\subsection{Design Challenges}

\begin{figure}[htbp]
  \centering
  % Requires \usepackage{graphicx}
  \includegraphics[width=0.5\textwidth]{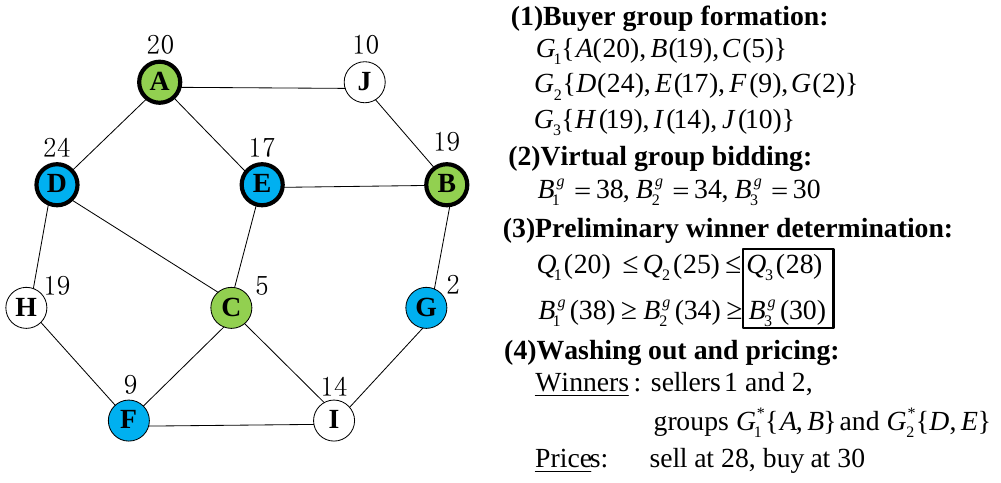}\\
  \caption{An example of TDSA auctions. There are 3 sellers and 10 buyers in a double spectrum auction. The requests of the three sellers are $Q_1 = 20$, $Q_2 = 25$, and $Q_3 = 28$, respectively. The conflict graph of buyers and the bids of the buyers are shown. Applying TDSA procedure, the auction outcome can be finally obtained.}\label{fig:example}
\end{figure}

Recently, a secure mechanism for double spectrum auctions based on TRUST, called PS-TRUST, has been proposed \cite{14chenz}. However, to achieve a better auction performance, a secure version of TDSA mechanism is demanded. We argue that secure spectrum auction design based on TDSA is challenging. With the example of TDSA auctions as depicted in Fig.~\ref{fig:example}, we show the design challenges as follows.

First, the direct application of the design rationale of PS-TRUST to TDSA makes the resulted mechanism insecure. Following the design of PS-TRUST, the step of buyer group formation is performed without any security guarantee by the auctioneer, who thus obtains all buyer groups $G_1$, $G_2$ and $G_3$. Additionally, the auctioneer certainly knows the auction outcome including winners (sellers 1 and 2, and buyer groups $G^*_1$ and $G^*_2$) and prices (selling price 28 and buying price 30) at the end of the auction. Therefore, the auctioneer can easily know which buyers in $G_1$ and $G_2$ are washed out, and their bid ranges, e.g., buyer $C$ is washed out from $G_1$ and its bid value is no more than $10$ ($=\frac{30}{3}$). However, these information goes beyond the auction outcome, and the security requirements are not satisfied. The security of this auction design fails.

Second, securing the step of buyer group formation completely will lead to a heavy performance overhead. One way to address the first challenge is to form buyer groups securely. However, a completely secure buyer group formation protocol will incur a very high computational complexity, e.g., the direct conversion from a buyer group formation algorithm to a garbled circuit requires a computational complexity of $O(N^3)$, where $N$ is the number of buyers, and the performance overhead will be unbearable for practical applications.

Finally, TDSA has a more complicated auction procedure than TRUST. For example, TDSA needs steps of virtual group bidding and washing out, respectively, before and after the step of preliminary winner determination. How to perform the auction computation securely requires careful design.

\subsection{Design Rationale}

Our main idea is to securely perform double spectrum auction of TDSA between the auctioneer and the agent, by leveraging the techniques of secure two-party computation. Secure two-party computation aims at jointly computing a function $f$ between two parties over their inputs while keep those inputs private. All that parties can learn is what they can learn from the output and their own respective inputs. In this work, we consider the auction computation as a function $f$ whose inputs are the requests of $M$ sellers and the bids of $N$ buyers. Although it is possible for the auctioneer, the sellers and the buyers to jointly and securely compute the auction without introducing the agent, it would lead to a heavy communication overhead among the participants, making the secure auction less practical.

Particularly, we address the first two challenges by providing a secure buyer grouping protocol with a tradeoff between security and efficiency, and address the third challenge using the technique of garbled circuits. More specifically, to design such a secure grouping protocol, we let the auctioneer encrypt all node information (i.e., buyer IDs and bids) of the conflict graph using our joint encryption based on Paillier cryptosystem, and outsource the grouping procedure to the agent, who obtains and randomizes all the groups and returns them to the auctioneer. This grouping protocol makes the auctioneer learn nothing about the grouping information (e.g., which buyers in which group), preventing it from learning which buyers are washed out and what their bid ranges are, and thus achieves higher security. To make use of garbled circuits, we design an efficient data-oblivious auction algorithm, whose execution path does not depend on its input, and then turn it into the boolean circuit that the garbled circuit is based on.

\section{SDSA: Design Details and Proofs}\label{sec:sdsa-details}
In this section, we elaborate on the design of SDSA, and prove that it is secure against semi-honest adversaries.

\subsection{Overall Protocol}

Overall, SDSA is a secure protocol for double spectrum auctions executed mainly by the auctioneer and the agent. To achieve security and efficiency, this protocol combines three security techniques, homomorphic encryption (i.e., Paillier cryptosystem), garbled circuits (a.k.a. Yao's protocol), and secret sharing. Specifically, homomorphic encryption is used for secure buyer grouping, garbled circuits are used for secure auction processing, while secret sharing is used for the switch between the former two techniques.

Assume that the auctioneer (A) and the agent (B) have public/private key pairs of Paillier cryptosystem, $(pk_{A}, sk_{A})$ and $(pk_{B}, sk_{B})$, respectively, and there are authenticated secure channels between the auctioneer and the agent, as well as between the auctioneer and each seller or each buyer. SDSA comprises three phases as follows. In Phase I, all sensitive inputs (i.e., all sellers' IDs and requests, and all buyers' IDs and bids) are encrypted with the agent's public key $pk_B$, and submitted to the auctioneer, while the insensitive inputs (all buyers' geographic locations) are submitted to the auctioneer in the clear. Note that in this phase the auctioneer can only know the insensitive inputs, but not the sensitive inputs. In Phase II, both the auctioneer and the agent cooperatively perform the secure group formation, such that anyone of them knows little about the group information. In Phase III, both the auctioneer and the agent cooperatively perform the secure auction and obtain the auction outcome, without leaking any information about the sensitive inputs beyond the auction outcome. The overall protocol is shown in Protocol~\ref{pro:secureauction}.

\floatname{algorithm}{Protocol}

\begin{algorithm}[htbp]
\caption{SDSA: Secure Double Spectrum Auction}
\label{pro:secureauction}
\begin{algorithmic}[1] % 这个1 表示每一行都显示数字
\Require  % 算法的输入参数：Input
The boolean auction circuit publicly known.
\Ensure  % 算法的输出：Output
Auction outcome in the clear.

%\noindent\textbf{Phase 1: Bid Submission and Bid Sharing}
%\rule{0.4\textwidth}{1pt}
%\noindent

\Algphase{Phase I: Submission}
\State \textbf{Sellers:} Each seller $m$ submits to the auctioneer a tuple $([ID^s_m]_{pk_{B}}, [Q_{m}]_{pk_{B}})$.
\State \textbf{Buyers:} Each buyer $n$ submits to the auctioneer a tuple $((x_{n}, y_{n}), [ID^b_n]_{pk_{B}}, [B_{n}]_{pk_{B}})$.

\Algphase{Phase II: Secure group formation}

\State \textbf{Auctioneer:} Constructs a conflict graph of buyers, and transforms each seller or buyer tuple as follows.
\[
\begin{split}
([ID^s_m]_{pk_{B}}, [Q_{m}]_{pk_{B}}) &\rightarrow ([\![ID^s_m]\!], [\![Q_{m}]\!])\\
([ID^b_n]_{pk_{B}}, [B_{n}]_{pk_{B}}) &\rightarrow ([\![ID^b_n]\!], [\![B_{n}]\!])
\end{split}
\]
Sends the conflict graph, all buyer tuples and the agent's corresponding parts of all seller tuples to the agent.

\State \textbf{Agent:} Performs bid-independently buyer group formation and gets a group set $G=\{G_{1},G_{2},\ldots,G_{T}\}$. Then SS-blinds the buyer tuples of each group $G_t$, pads $G_t$ to $R = \max_t R_t$ buyer tuples, and finally returns the SS-blinded and padded group set $G^*$ to the auctioneer.

\Algphase{Phase III: Garbled circuit computing}

\State \textbf{Agent:} Decrypts its corresponding parts of both group set $G^*$ and all seller tuples to get its share of sensitive inputs, and garbles the auction circuit and its share of sensitive inputs, and sends the garbled circuit, the garbled values obtained and output decoding to the auctioneer.
\State \textbf{Auctioneer:} Decrypts its corresponding parts of both group set $G^*$ and all seller tuples to get its share of sensitive inputs, and executes the garbled circuit to compute the auction outcome in the clear, with the agent's garbled share and its plain share of sensitive inputs, and the output decoding.
%\RETURN $\cdots$; % 算法的返回值
\end{algorithmic}
\end{algorithm}

\subsection{Detailed Design}

Now we present the detailed design of each phase as follows.

\textbf{Phase I: Submission}

In this phase, sellers and buyers provide the auctioneer with inputs of the auction. There are two types of inputs, sensitive and insensitive, according to the relevance to requests or bids (sellers' or buyers' privacy). In light of this, sensitive inputs include all sellers' IDs and requests, and all buyers' IDs and bids, while insensitive inputs include all buyers' locations. All sensitive inputs are encrypted with the agent's public key $pk_B$ and submitted to the auctioneer, preventing the auctioneer from learning them, while insensitive inputs are submitted in the clear also to the auctioneer. A seller or a buyer can go off-line after submission. The tuples submitted by the sellers and buyers are described as follows.
\begin{description}
  \item [Seller] $m$: $([ID^s_m]_{pk_{B}}, [Q_{m}]_{pk_{B}})$, for $m=1,2,...,M$
  \item [Buyer] $n$: $((x_{n}, y_{n}), [ID^b_n]_{pk_{B}}, [B_{n}]_{pk_{B}})$, for $n=1,2,...,N$
\end{description}

Since we have assumed that all communication channels are authenticated secure, all sensitive inputs would keep private and would not be eavesdropped. Although the auctioneer obtains all the inputs needed for the auction, it knows nothing about the sensitive inputs.

\textbf{Phase II: Secure group formation}

In this phase, the auctioneer first constructs a conflict graph of buyers according to their geographic locations. Secondly, the auctioneer reforms the seller or buyer tuples as follows.
\begin{description}
  \item [Seller] $m$: $([ID^s_m]_{pk_{B}}, [Q_{m}]_{pk_{B}})$, for $m=1,2,...,M$
  \item [Buyer] $n$: $([ID^b_n]_{pk_{B}}, [B_{n}]_{pk_{B}})$, for $n=1,2,...,N$
\end{description}
These tuples are then converted into the joint encryption forms between the auctioneer and the agent as described in Section~\ref{sec:preliminaries:jointencryption}:
\begin{description}
  \item [Seller] $m$: $([\![ID^s_m]\!], [\![Q_{m}]\!])$, for $m=1,2,...,M$
  \item [Buyer] $n$: $([\![ID^b_n]\!], [\![B_{n}]\!])$, for $n=1,2,...,N$
\end{description}

Finally, the auctioneer sends the conflict graph, all buyer tuples $\{([\![ID^b_n]\!], [\![B_{n}]\!])\}_{n=1}^N$, and the agent's corresponding parts of all seller tuples $\{([\![ID^s_m]\!]_B, [\![Q_{m}]\!]_B)\}_{m=1}^M$ to the agent. Note that here the conflict graph is ``blinded'', i.e., it merely indicates whether one buyer conflicts with another buyer, but all buyers are anonymous.

Upon receiving the conflict graph from the auctioneer, the agent executes a grouping algorithm to divide the buyers into several groups, where the members of the same group will not conflict with each other when using an identical channel. Notice that this grouping operation is bid-independent, and we denote the set of non-conflict buyer groups by $G=\{G_{1},G_{2},\ldots,G_{T}\}$, with $R_{t}$ buyer tuples in each group $G_{t}$.

Next, the agent blinds the buyer tuples of each group $G_t$ using the SS-blinding of the joint encryption, and pads each group $G_t$ with null buyers $([\![0]\!], [\![0]\!])$ to $R = \max_t R_t$ buyer tuples.
This blinding prevents the auctioneer from identifying buyers by matching the values of its parts in the jointly encrypted buyer tuples, while the padding hides the difference of buyer groups in member size. Afterward, the agent returns these SS-blinded and padded group set $G^*$ to the auctioneer.

\textbf{Phase III: Garbled circuit computing}

In this phase, both the auctioneer and the agent hold their corresponding parts of jointly encrypted group set $G^*$ and seller tuples. Each of them decrypts its respective parts to get its share of all sensitive inputs necessary for the auction. Then, the agent generates a garbled circuit with shared inputs based on the given auction Boolean circuit, garbles its share of sensitive inputs, and sends the garbled circuit, the garbled values, and the decoding table for the outcome to the auctioneer. Finally, the auctioneer evaluates the garbled circuit and decodes the output.

For this phase, the most important thing is to design a data-oblivious algorithm based on TDSA auction mechanism. The data-oblivious algorithm can then be easily converted into a Boolean circuit, which is used in this phase. We describe this data-oblivious algorithm shown in Algorithm~\ref{alg:SDSAauction} as follows.

\floatname{algorithm}{Algorithm}
\begin{algorithm}[!htb]\small
\caption{Data-oblivious SDSA Auction}
\label{alg:SDSAauction}
\begin{algorithmic}[1] % 这个1 表示每一行都显示数字
\Require  % 算法的输入参数：Input
The set of sellers $S$, the set of buyer groups $G^*$
\Ensure % 算法的输出：Output
Winners and their clearing prices

\Algphase{\emph{(1) Initialization:}}
\State Initialize matrix $\mathbb{S}$, $\mathbb{G}$, $\mathbb{G}_{t}$
\label{code:auction:Initialize}% 对此行的标记，方便在文中引用算法的某个步骤

\Algphase{\emph{(2) Secure virtual group bidding:}}

%\STATE Compute the group bidding in $\mathbb{G}$:
\For {$t=1 \ldots T$}
\label{code:auction:Compute GBid}

    \State Sort $\mathbb{G}_{t}$ in non-ascending order of $B_{t,j}$
    \State //Abuse same indexes after sorting
    \label{code:auction:Sort buyers}

    %\STATE Enumerate the virtual group bids in $\mathbb{G}_{t}$:

    \For {$j=1 \ldots R$}
        \State $B^{v}_{t,j}\leftarrow B_{t,j} \times j$
        \State $B^{g}_{t}\leftarrow max(B^{g}_{t}, B^{v}_{t,j})$
    \EndFor
    \label{code:auction:enumerate $B^{v}_{j}$}

\EndFor

\Algphase{\emph{(3) Secure preliminary winner determination:}}

\State Sort $\mathbb{S}$ in non-descending order of $Q_i$\label{code:auction:Sort sellers}
\State Sort $\{B_t^g\}_{t=1}^T$ in non-ascending order \label{code:auction:Sort groups}

\State $\mathbbm{m} \leftarrow min(T,M)$
\For{$\phi=\mathbbm{m} \ldots 2$}
\State $\lambda_{\phi}\leftarrow(B^{g}_{\phi} \geq Q_{\phi} \wedge B^{g}_{\phi} \neq B^{g}_{\phi-1})$ \label{line:McAfee-rule1}
\EndFor
%\STATE $\lambda_{1}\leftarrow\lambda_{2}$

\State $\delta_{\mathbbm{m}}\leftarrow \lambda_{\mathbbm{m}}$ \label{line:delta1}
\State $P^{s} \leftarrow Q_{\mathbbm{m}} \cdot \delta_{\mathbbm{m}}$ \label{line:seller1}
\State $P^{g} \leftarrow B_{\mathbbm{m}}^{g} \cdot \delta_{\mathbbm{m}}$ \label{line:group1}
%\State $\{w_{k}^{s}\}_{k=\Phi}^{M}\leftarrow\{0\}_{M-\Phi+1}$
%\STATE $\{w_{k}^{g}\}_{k=min\{T,M\}}^{T}\leftarrow\{0\}_{T-min\{T,M\}}$
\For{$\phi= \mathbbm{m}-1 \ldots 1$}
\State $\lambda_{\phi}\leftarrow \lambda_{\phi}\vee\lambda_{\phi+1}$ \label{line:McAfee-rule2}
\State $\delta_{\phi}\leftarrow \lambda_{\phi}\oplus \lambda_{\phi+1}$ \label{line:delta2}
\State $P^{s} \leftarrow P^{s}+Q_{\phi} \cdot \delta_{\phi}$ \label{line:seller2}
\State $P^{g} \leftarrow P^{g}+B_{\phi}^{g} \cdot \delta_{\phi}$ \label{line:group2}
\State $w^{s}_{\phi}\leftarrow \lambda_{\phi+1}$ \label{line:sellerwinner}
\EndFor

\Algphase{\emph{(4) Secure washing out:}}

\For{$t=1 \ldots T$}
\State $w^{g}_{t}\leftarrow(B^{g}_{t}>P^{g})$ \label{line:groupwinner}
\State $w^{b}_{t,R}\leftarrow w^{g}_{t} \wedge (B^{v}_{t,R} > P^{g})$ \label{line:washing1}

    \For{$j=R-1 \ldots 1$ }
    \State $w^{b}_{t,j}\leftarrow w^{g}_{t} \wedge ((B^{v}_{t,j} > P^{g}) \vee w^{b}_{t,j+1})$ \label{line:washing2}
    \State $C_{t}\leftarrow max(C_{t}, j\cdot(w^{g}_{t}\wedge w^{b}_{t,j}))$ \label{line:ct}
    \EndFor
    \State $P^{g}_{t}\leftarrow w^{g}_{t}\cdot (P^{g}/C_{t})$ \label{line:pgt}
\EndFor

\Algphase{\emph{(5) Pricing:}}

\State Collect clearing prices for sellers: $\mathcal{S} = \{(ID^s_i,P^{s}_{i})\}_{i=1}^M$
\State Collect clearing prices for buyers: $\mathcal{B} = \{(ID^b_j,P^{b}_{j})\}_{j=1}^N$
\State Sort $\mathcal{S}$ in term of $ID^s_i$, and Sort $\mathcal{B}$ in term of $ID^b_j$\\
\Return $\mathcal{S}$ and $\mathcal{B}$ % 算法的返回值
\end{algorithmic}
\end{algorithm}

\textbf{\emph{Step 1: Initialization}}

In this step, two arrays $\mathbb{S}$ and $\mathbb{G}$ are constructed to represent all sellers and all buyer groups, and an array $\mathbb{G}_{t}$ is constructed to represent all buyers of each group $t$. In these arrays, each seller, each group, and each buyer are represented by a seller tuple, a group tuple, and a buyer tuple, respectively:
\begin{description}{\desclabelstyle{\pushlabel}\desclabelwidth{6em}}
  \item [Seller] tuples: $(\emph{ID}_i, Q_{i}, w^{s}_{i})$, $i\in [1..M]$
  \item [Group] tuples: $(B^{g}_{t}, w^{g}_{t}, C_{t}, P^{g}_{t})$, $t\in [1..T]$
  \item [Buyer] tuples: $(\emph{ID}_{t,j}, B_{t,j}, B^{v}_{t,j}, w^{b}_{t,j})$, $j\in [1..R]$, $t\in [1..T]$
\end{description}
where $R = \max_t R_{t}$, and $w^{s}_{i}$, $w^{g}_{t}$, and $w^{b}_{j}$ are binary flags indicating whether seller $i$, group $t$, and buyer $j$ is a winner ($1$) or not ($0$), respectively.

The resulted arrays $\mathbb{S}$, $\mathbb{G}$, and $\mathbb{G}_t$ are initialized as follows, where the ``null" symbol $\bot$ is a placeholder.
\begin{displaymath}\footnotesize
\mathbb{S}=
\left( \begin{array}{rccc}
i: & 1 & \cdots & M \\
\emph{ID}_i:& \emph{ID}_1 & \cdots & \emph{ID}_M \\
Q_i:& Q_1 & \cdots & Q_M \\
w_i^s:& 0 & \cdots & 0 \\
\end{array} \right),
\mathbb{G}=
\left( \begin{array}{rccc}
t: & 1 & \cdots & T \\
B_t^g:& \bot & \cdots & \bot \\
w_t^g:& 0 & \cdots & 0 \\
C_t:& 0 & \cdots & 0 \\
P_t^g:& \bot & \cdots & \bot \\
\end{array} \right)
\end{displaymath}
\vspace{0.2cm}
\begin{displaymath}\footnotesize
\mathbb{G}_t=
\left( \begin{array}{rccc}
j: & 1 & \cdots & R \\
\emph{ID}_{t,j}:& \emph{ID}_{t,1} & \cdots & \emph{ID}_{t,R} \\
B_{t,j}:& B_{t,1} & \cdots & B_{t,R} \\
B_{t,j}^v:& \bot & \cdots & \bot \\
w_{t,j}^b:& 0 & \cdots & 0 \\
\end{array} \right)
\end{displaymath}

\textbf{\emph{Step 2: Secure virtual group bidding}}

This step computes virtual group bids $B^{v}_{t,j}$ and group bids $B^{g}_t$ with two for-loops. The outer for-loop iterates over the groups, while the nested for-loop computes the group bid $B^{g}_{t}$ for each group $\mathbb{G}_{t}$:

1) The outer for-loop sorts buyers of $\mathbb{G}_{t}$ in non-ascending order of $B_{t,j}$, and the buyers will be deleted in this order at ``Secure washing out'' step.

2) Then, the nested for-loop enumerates all virtual group bids $B^{v}_{t,j}$ of $\mathbb{G}_{t}$, and selects the highest virtual group bid as the group bid of group $\mathbb{G}_{t}$.

\textbf{\emph{Step 3: Secure preliminary winner determination}}

This step employs McAfee's design to preliminarily determine winning sellers and winning groups. Since this procedure contains many operations such as comparisons and selections depending on requests or bids, it is challenging to convert it into a data-oblivious algorithm. To address the challenge, a general solution is to introduce appropriate binary flags indicating various conditions, and then use them to data-obliviously determine critical sellers and groups, and finally compute the corresponding clearing prices. This is done via the following steps.

First, sellers are sorted in non-descending order of requests, and group bids are sorted in non-ascending order. Notice that here group bids rather than group tuples are sorted, so that only values other than tuples are swapped during the sorting, and this reduces computation overhead. Whether a group is a winning group can be determined simply in ``Secure washing out'' step by comparing it's group bid with the clearing price $P^{g}$ for groups.

Second, in order to find out the critical index $\Phi$ data-obliviously, two arrays of binary flags, $\lambda_{\phi}$ and $\delta_{\phi}$ with $\phi\in\{1,...,\mathbbm{m}\}$, where $\mathbbm{m} = \min{(T, M)}$, are needed and work as follows:

$\lambda_{\phi}$: indicates whether $\phi$ is less than or equal to the critical index $\Phi$ in McAfee rules ($\lambda_{\phi}=1$) or not ($\lambda_{\phi}=0$) (Line~\ref{line:McAfee-rule1} \& \ref{line:McAfee-rule2}).

$\delta_{\phi}$: indicates whether $\phi$ is equal to the critical index $\Phi$ in McAfee rules ($\delta_{\phi}=1$) or not ($\delta_{\phi}=0$) (Line~\ref{line:delta1} \& \ref{line:delta2}).

According to the auction logic, the two flag arrays should take values with the following pattern:

\begin{center}
$\left(
\begin{array}{rccccccc}
   \phi: & 1 & \cdots & \Phi-1 & \Phi & \Phi+1 & \cdots & \mathbbm{m} \\
    \lambda_{\phi}: & 1 & \cdots & 1 & 1 & 0 & \cdots & 0 \\
    \delta_{\phi}: & 0 & \cdots & 0 & 1 & 0 & \cdots & 0 \\
  \end{array}
\right)$
\end{center}

Then, it is straightforward to compute the price for $\Phi-1$ winning sellers and $\Phi-1$ winning groups (Line~\ref{line:seller1} \& \ref{line:seller2}, Line~\ref{line:group1} \& \ref{line:group2}):
\begin{center}
$P^{s} = P^{s}+Q_{\phi} \cdot \delta_{\phi}$, $P^{g} = P^{g}+B^{g}_{\phi} \cdot \delta_{\phi}$.
\end{center}

In the end, binary flag $w^{s}_{\phi}$ of seller $\phi$ can be updated as follows: when $k\geq \mathbbm{m}$, seller $\phi$ is bound to be a loser, and it remains that $w_{\phi}^{s}=0$ as initialization; when $1\leq \phi\leq \mathbbm{m}-1$, $w^{s}_{\phi}$ can be determined by $w^{s}_{\phi}=\lambda_{\phi+1}$ (Line~\ref{line:sellerwinner}).

\textbf{\emph{Step 4: Secure washing out}}

This step eliminates buyers among winning buyer groups, whose bids are too low to afford the clearing price $P^g$. To achieve data-obliviousness, we have to do the same washing out for every group, and use $w^{b}_{t,j}$ to mark whether buyer $j$ of group $\mathbb{G}_t$ is a final winning buyer. This is done as below.

Firstly, the outer for-loop iterates over all buyer groups, and sets values to binary flags $w^{g}_{t}$ by comparing group bids with the clearing price $P^g$ for groups (Line~\ref{line:groupwinner}).

Secondly, an inner for-loop selects the largest virtual group of $G_t$, whose virtual bid is larger than $P^{g}$, and deletes other buyers outside this largest virtual group (Line~\ref{line:washing1} \& \ref{line:washing2}). Meanwhile, the number of winning buyers in group $t$ (i.e. $C_{t}$) is computed (Line~\ref{line:ct}).

Finally, the price for each winning group $t$ is shared by its winning members: $P^{g}_{t}=P^{g}/C_{t}$ (Line~\ref{line:pgt}).

%Similar to the last/previous step, the auctioneer first blinds and sends the winning groups to the agent and subsequently the agent prepares a Yao's garbled circuit as followed for each winning groups:
%
%(a) takes as input the garbled values corresponding to the part of the buyer's \emph{ID}s and bids which are encrypted by the auctioneer;
%
%(b) recovers the true values corresponding to the buyer's \emph{ID} and bids by adding the values shared between the auctioneer and the agent;
%
%(c) performs the washing out procedure in the TDSA;
%
%(d) outputs the winning auction buyers and their clearing price in each winning groups respectively.
%
%Then, the agent sends those garbled circuits to the auctioneer. Next, the agent enroll in an oblivious transfer protocol with the auctioneer so that the auctioneer obtains garbled values corresponding to the part of buyer's \emph{ID}s and bids encrypted by the auctioneer. Finally, the auctioneer evaluates the circuit and gets the output.

\textbf{\emph{Step 5: Pricing}}

This step computes the prices for all sellers and all buyers. It first constructs an array $\mathcal{S}$ of $M$ seller tuples and an array $\mathcal{B}$ of $N$ buyer tuples. For each seller $i$, it creates a tuple $(ID^s_i,P^{s}_{i})$; For each buyer $j$, it creates a tuple $(ID^b_j,P^{b}_{j})$. Then, it extracts the IDs and clearing prices for buyers and sellers from the processed arrays $\mathbb{S}$, $\mathbb{G}$, and $\mathbb{G}_t$ with $t \in [1..T]$. Specifically, the clearing price for seller $i$ can be obtained by computing $P^{s}_{i}=w^{s}_{i}\cdot P^{s}$, the clearing price for buyer $j$ within group $t$ can be obtained by computing $P^{b}_{j}=w^{g}_{t}\cdot w^{b}_{j}\cdot P^{g}_{t}$. In order to prevent an adversary from inferring the request ranking or bid ranking information based on the order of sellers or buyers in the auction outcome, the algorithm should reorder the winning sellers or the winning buyers in term of \emph{ID} or any other attribute independent of requests or bids before outputting the auction outcome.
%the announcement of auction outcome.

\subsection{Security Analysis}

We now prove the security of Protocol~\ref{pro:secureauction} with the security requirements and security definition described in Section~\ref{sec:security-req}. As stated in the same section, it is easy to verify that each seller, or each buyer, or anyone not participating in the auction learns nothing about the sensitive inputs except what can be revealed from the auction outcome. Thus, we mainly prove that through running the proposed protocol neither the auctioneer nor the agent can learn anything about the sensitive inputs beyond what is revealed by the auction outcome, and this can be stated as Theorem \ref{the:semi-honest}.

\begin{mytheorem}\label{the:semi-honest}
As long as the auctioneer and the agent are not corrupted simultaneously, Protocol \ref{pro:secureauction} is secure against semi-honest adversaries.
\end{mytheorem}

\textbf{Proof:} To prove the security of Protocol \ref{pro:secureauction}, we first prove the security of each phase in two separate
cases, depending on which party is corrupted. For each phase, we show that for all PPT adversaries, the
corrupted party's view based on the auctioneer and the agent's interaction is indistinguishable
to its view when it interacts with a simulator instead. In other words, we show
that there exist simulators $S1$ and $S2$ that satisfy conditions
\eqref{equ:semisecure1} and \eqref{equ:semisecure2} for each phase.
Then, due to the sequential composition theorem \cite{10Hazayc}, we actually prove the security of the entire protocol.

\textbf{Phase I Submission:} In this phase, each seller (resp. buyer) encrypts its ID and request (resp. bid) with the agent's public key $pk_B$, and submits these encrypted sensitive inputs to the auctioneer. Since the encryption (i.e., Paillier encryption) is semantically secure, the auctioneer cannot decrypt the ciphertexts. Also, each buyer sends its geographic location to the auctioneer, but location information is supposed to be input of the auctioneer, and is independent on the sensitive inputs. Thus, the auctioneer learns nothing about the sensitive inputs. Meanwhile, the agent sees nothing in this phase, and it certainly learns nothing about the sensitive inputs. Therefore, this phase is secure.

\textbf{Phase II Secure group formation:} In this phase, the auctioneer's view includes:
\begin{itemize}
\item The view (e.g., the conflict graph of buyers, the joint ciphertexts) that can be inferred from the ciphertexts and buyers' locations just as it sees in Phase I;
\item $T$ buyer groups with $R$ members each, with no IDs or bids known.
\end{itemize}
The agent's view includes a share of all sellers' IDs and requests, and that each buyer conflicts with some other buyers, and does not conflict with the remaining other buyers, with no IDs and bids known.

It is straightforward to verify that both views are independent on the sensitive inputs (i.e., IDs, requests, and bids), and neither the auctioneer or the agent can learn anything about the sensitive inputs. Thus, this phase is secure.

\textbf{Phase III Garbled circuit computing:} In this phase, based on a data-oblivious auction algorithm, which can be easily converted into a Boolean circuit, the agent constructs a garbled circuit and garbles its shares of the sensitive inputs. The agent then sends the garbled circuit, garbled values and decoding information to the auctioneer. Upon receiving all the information sent from the agent, the auctioneer uses them together with its shares of the sensitive inputs to compute the auction outcome in the clear. This phase is a classical application of the garbled circuit protocol (a.k.a. Yao's protocol), and its security has been proved in \cite{Lindell2009APO}.

Since the above three phases are composed sequentially,  it follows from the
sequential composition theory that Protocol~\ref{pro:secureauction}
is secure against semi-honest adversaries. $\Box$

\section{Performance Evaluation}\label{sec:experiment}

In this section, we evaluate the performance of SDSA in term of computation and communication overheads, and compare it with the original auction mechanism TDSA.

\subsection{Experimental Method}
Our simulation experiments are conducted using Java on a 64-bit Windows 7 desktop with Intel(R) core(TM) i5 CPU @ 3.30GHz and 8GB of memory. All experimental results are averaged over 10 runs. Especially, we implement SDSA with Paillier cryptosystem with 1024-bit plaintext space and FastGC, a garbled circuit computation framework. The auctioneer and the agent are simulated by two processes on the machine.

To evaluate SDSA extensively, different experimental parameters including the number of buyers, the number of sellers, and the bit length of requests or bids, are configured. Usually, for an experiment, one of the above parameters varies, while the other two are fixed. For the sake of brevity, we only present the default configuration here. Buyers are randomly distributed in a $2000\text{m} \times 2000\text{m}$ area, and the interference range is $500$m. The requests of sellers are uniformly distributed in $[1,151)$, while the bids of buyers are uniformly distributed in $[1,51)$. Requests, bids, and IDs are represented in 16 bits by default. For the joint encryption, we by default let $K = 32$ (Note that we combine an ID and a bid or request as a value), and $K^{\prime} = 81$.

%As for the group formation, which is equivalent to finding independent set problem, we adopt existing algorithm based on Branch and Bound Method to simulate it.

\subsection{Experimental Results}

We conduct experiments to compare the performance of SDSA and TDSA as: (1) the number of buyers varies; (2) the number of sellers varies. We also conduct an experiment to evaluate the performance of SDSA as the bit length of requests or bids varies.

\textbf{(1) Number of buyers varies}

\begin{figure}[htbp]
  \centering
  \subfigure[Computation Overhead]{
    \label{fig:Buyer_N:a}
    \includegraphics[width=0.48\linewidth]{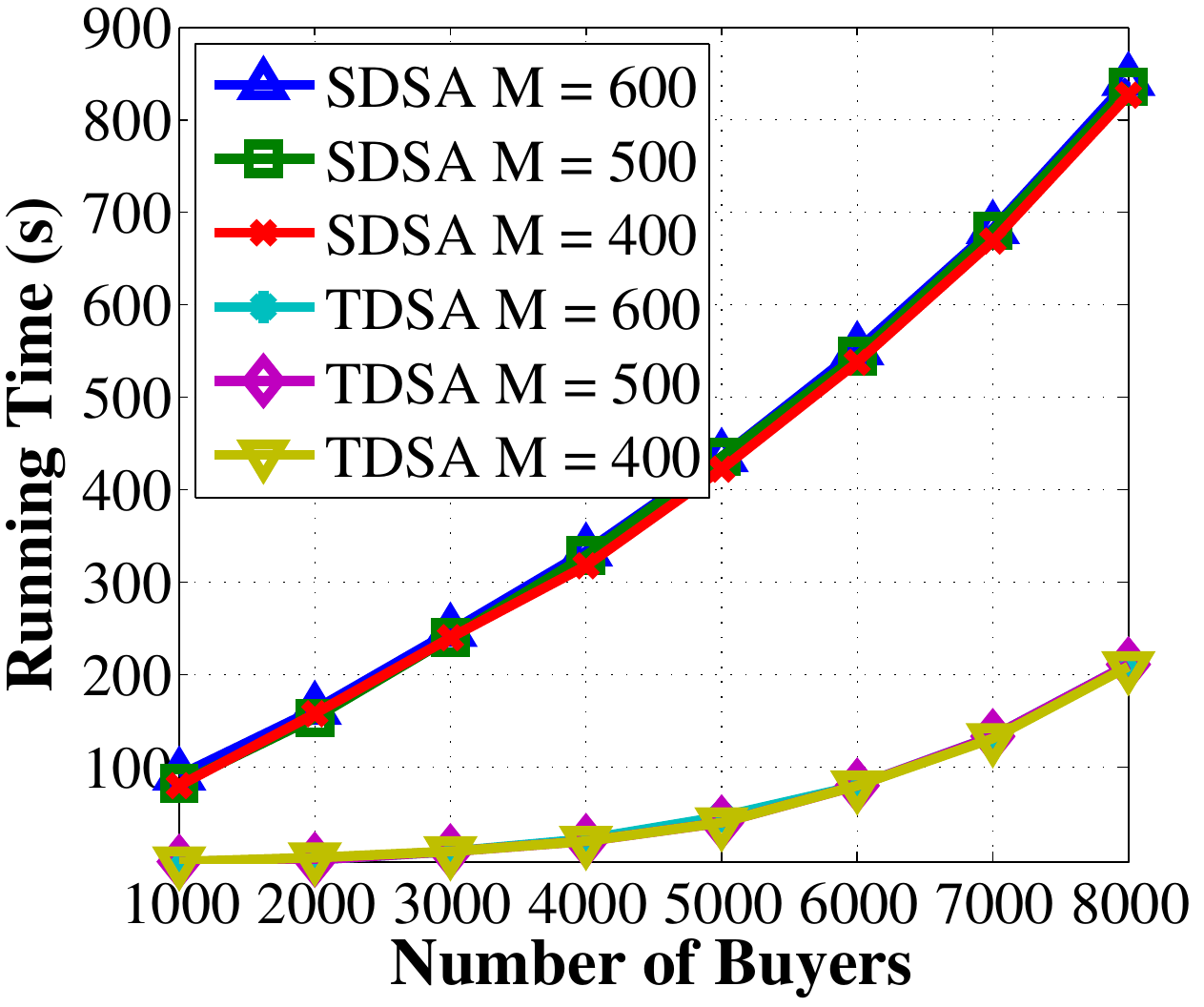}}
  \subfigure[Communication Overhead]{
    \label{fig:Buyer_N:b}
    \includegraphics[width=0.48\linewidth]{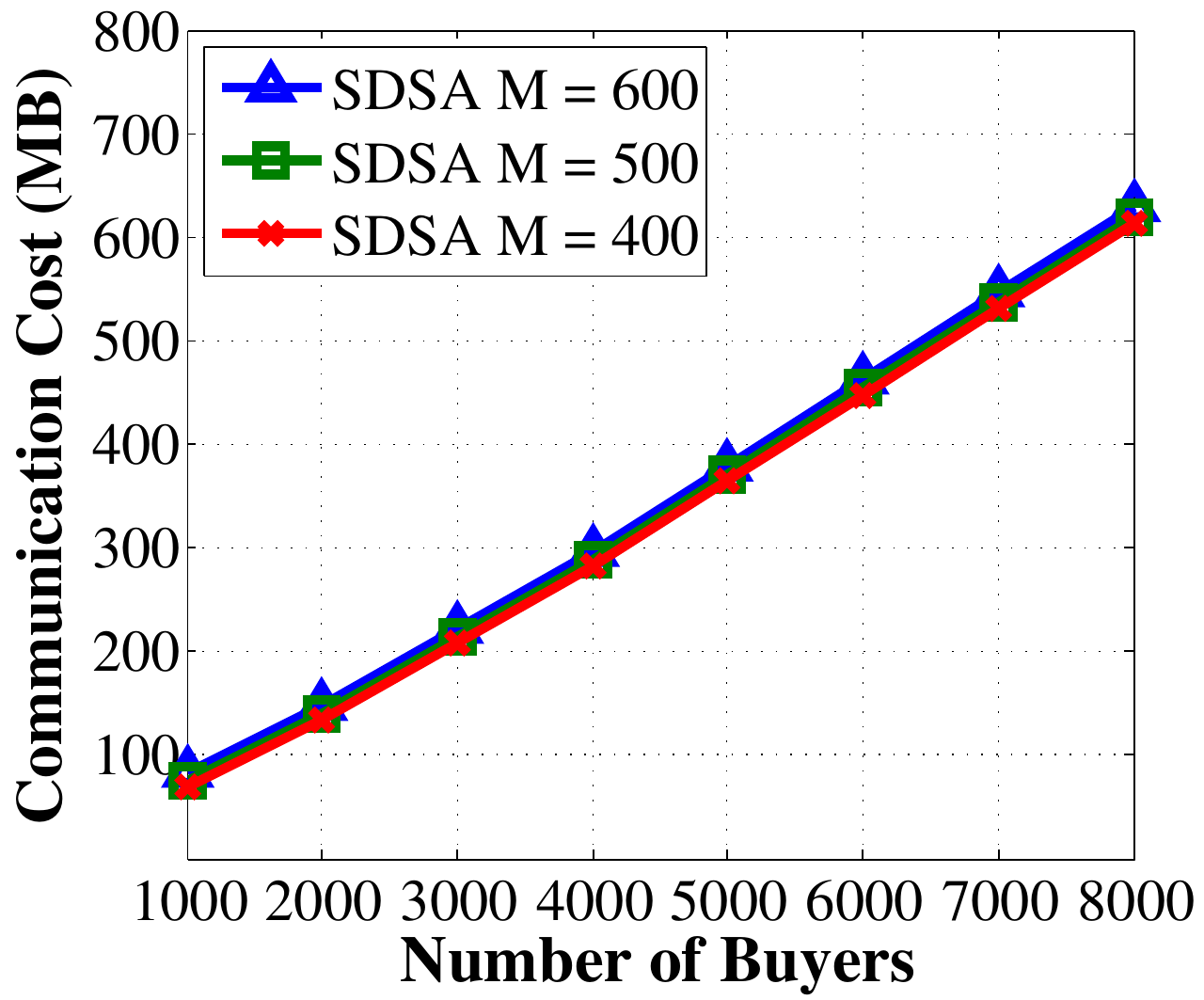}}
  %\hspace{1in}
  \caption{Computation and communication comparisons for our protocols as the number of buyers N varies}
  \label{fig:Buyer_N}
\end{figure}

Fig. \ref{fig:Buyer_N} plots the performance comparison between SDSA and TDSA, as the number of buyers $N$ increases from 1000 to 8000, and the number of sellers is fixed at $M=600$, $M=500$, and $M=400$. Specifically, Fig. \ref{fig:Buyer_N:a} shows that the running times of SDSA and TDSA follow the same growth pattern, which is approximately a quadratic polynomial of $N$. Furthermore, the gaps of running times between the two auctions are mild, e.g., when $M=600$, and $N = 7000$, the corresponding running times are 680s and 125s, respectively. Fig. \ref{fig:Buyer_N:b} shows that the communication cost of SDSA is nearly linear with $N$ (Note that TDSA is run merely by the auctioneer, and needs no communications). Also, both Figs. \ref{fig:Buyer_N:a} and \ref{fig:Buyer_N:b} illustrate that all curves with different $M$ values nearly overlap.

From the above observations, we can conclude that: (i) when $N$ is large, the computation overheads of both SDSA and TDSA tend to be dominated by the group formation algorithm, which consumes approximately quadratic polynomial time in $N$; (ii) SDSA incur limited computation and communication overheads compared to TDSA, and the relative overheads decrease as $N$ increases; (iii) the number of sellers $M$ has little effect on the computation and communication overheads, when $M$ is much smaller than $N$.

\textbf{(2) Number of seller varies}

\begin{figure}[htbp]
  \centering
  \subfigure[Computation Overhead]{
    \label{fig:Seller_M:a}
    \includegraphics[width=0.48\linewidth]{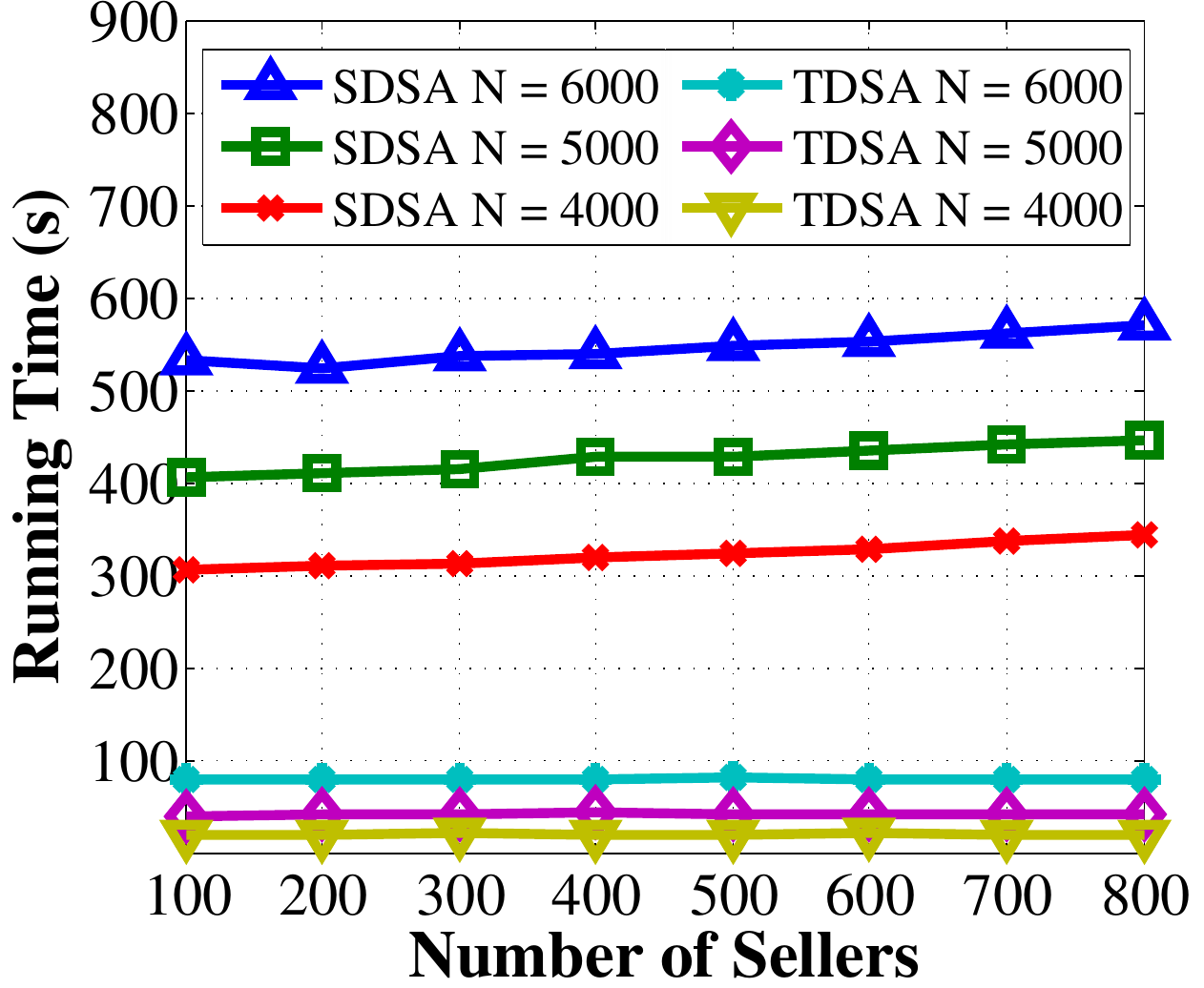}}
  \subfigure[Communication Overhead]{
    \label{fig:Seller_M:b}
    \includegraphics[width=0.48\linewidth]{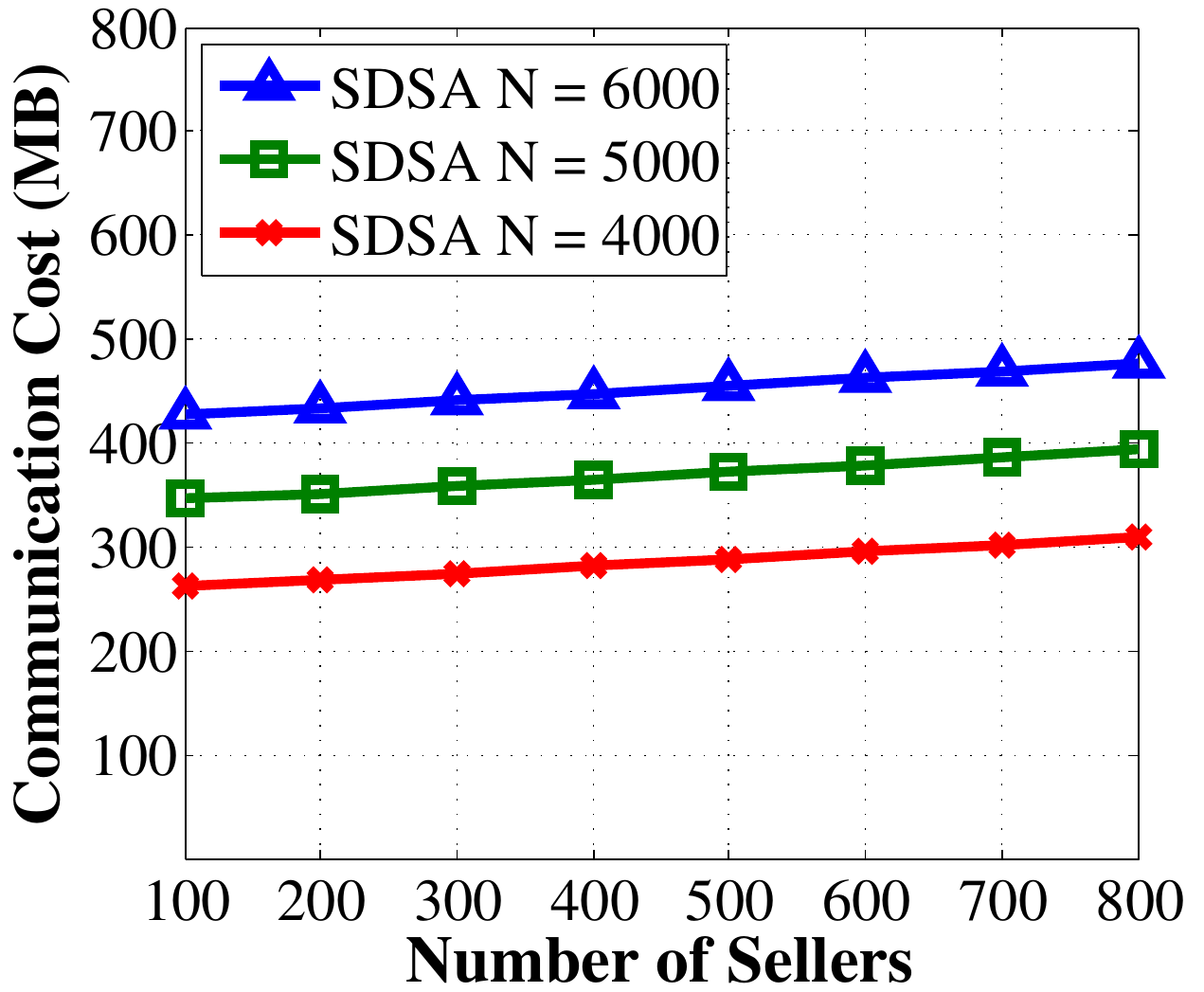}}
  %\hspace{1in}
  \caption{Computation and communication comparisons for our protocols as the number of sellers M varies}
  \label{fig:Seller_M}
\end{figure}

Fig. \ref{fig:Seller_M} depicts the performance comparison between SDSA and TDSA, as the number of sellers $M$ increases from 100 to 800, and the number of buyers $N=6000$, $N=5000$, and $N=4000$, respectively. The curves in both Figs. \ref{fig:Seller_M:a} and \ref{fig:Seller_M:b} are nearly flat, which validates point (iii) concluded above. Nonetheless, it looks like that $M$ has a stronger influence on the communication overhead than on the computation overhead. This is because the increase of sellers only adds some relatively few time-consuming calculations, and yet it does add some cipher data on the communication between the auctioneer and the agent.

\textbf{(3) Bit length varies}

\begin{figure}[htbp]\small
  \centering
  \subfigure[Computation Overhead]{
    \label{fig:l_bits:a}
    \includegraphics[width=0.48\linewidth]{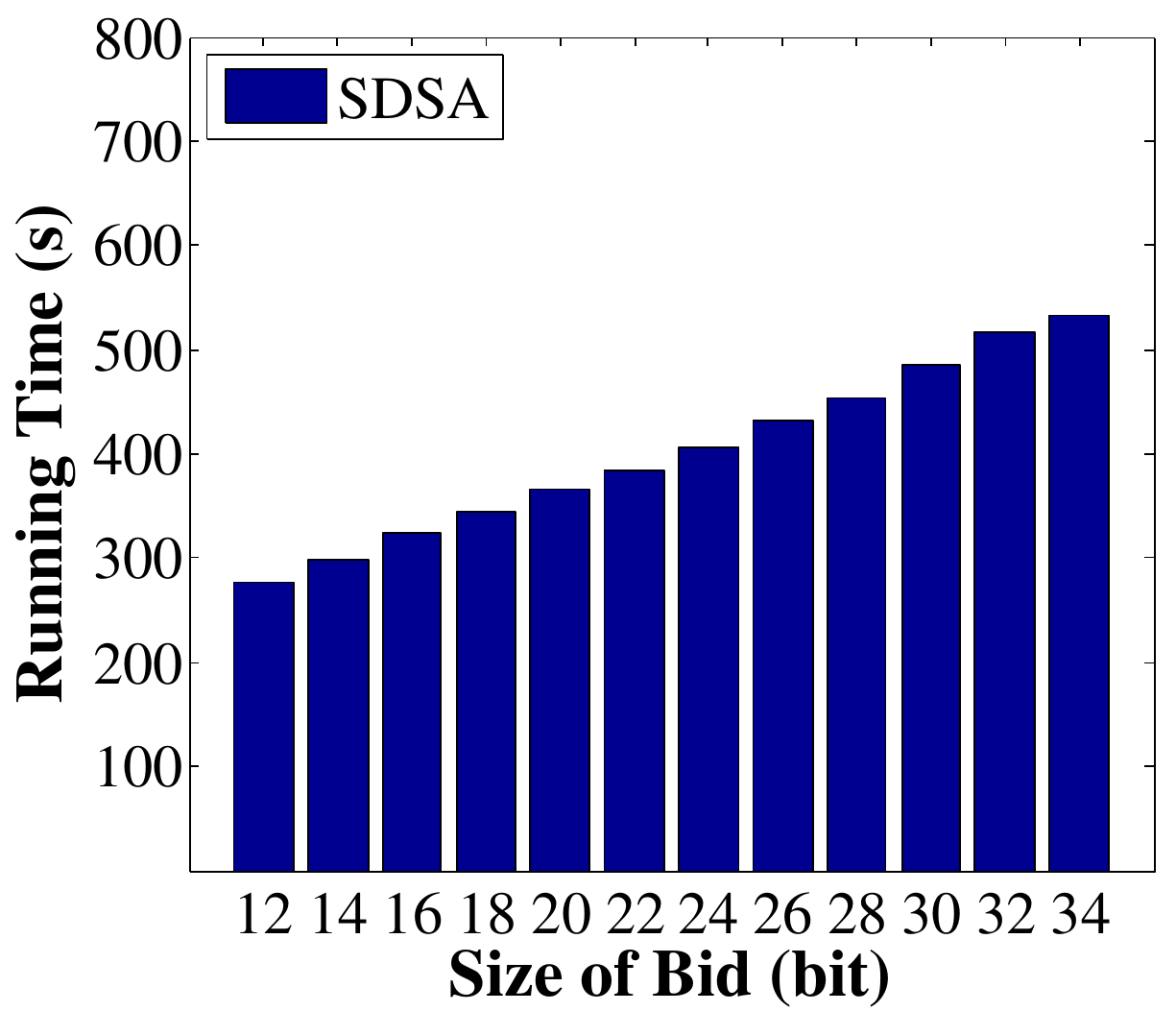}}
  \subfigure[Communication Overhead]{
    \label{fig:l_bits:b}
    \includegraphics[width=0.48\linewidth]{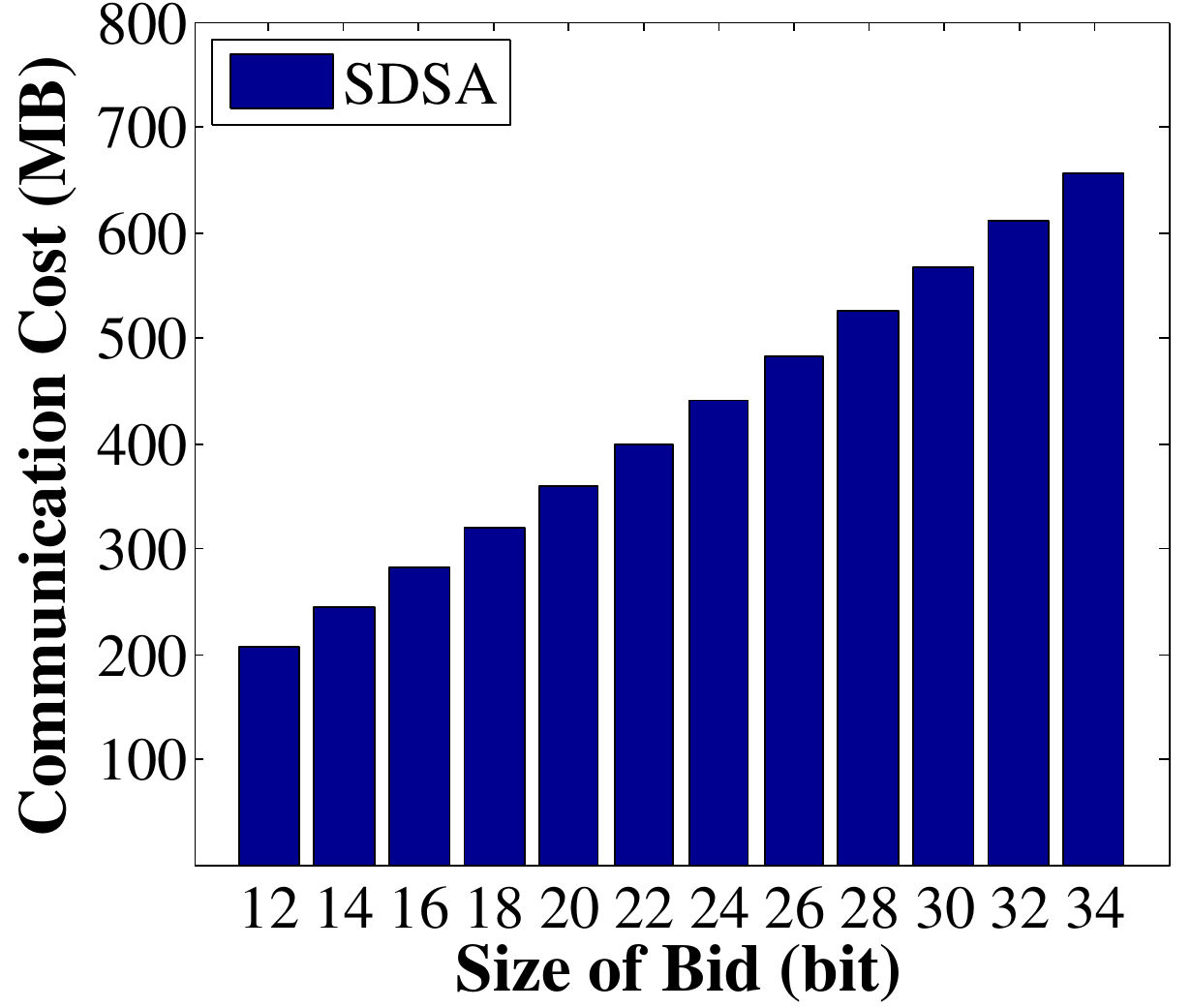}}
  %\hspace{1in}
  \caption{Computation and communication comparisons for our protocols as the bit length of bids varies}
  \label{fig:l_bits}
\end{figure}

In Fig.~\ref{fig:l_bits}, we vary the bit length of requests or bids from 12 to 34, while fix the number of buyers $N=4000$, and the number of sellers $M=400$. Both computation and communication overheads increase almost linearly with the bit length. The underlying reason is that each bit for every request or every bid need to be represented as a random key in garbled circuits, and the increase of bit length will undoubtedly raise the cost of both computation and communication.

From the above, we can conclude that SDSA is efficient in both computation and communication performance, even in large-scale double spectrum auctions, e.g., when there are 8000 buyers and 600 sellers, SDSA takes about 14 minutes time, and communicates about 600 MB data on a normal PC, and thus it is practical for real applications.

\section{Conclusion}\label{sec:conclusion}
In this paper, we have proposed SDSA, a privacy-preserving mechanism for double spectrum auctions. Basing on TDSA, SDSA provides a socially efficient spectrum auction. However, this in turn poses challenges for privacy achievement. We have tackled the challenges by leveraging on three security techniques: homomorphic encryption, secret sharing and garbled circuits. Specifically, we have designed a novel joint encryption based on homomorphic encryption to address the secure group formation, and have designed a data-oblivious auction algorithm and applied garbled circuits to address the secure auction procedure. Furthermore, we have implemented SDSA, and performed extensive experiments to illustrate that SDSA is efficient in performance and practical for real applications.

% conference papers do not normally have an appendix

% use section* for acknowledgement
%\section*{Acknowledgment}
%
%%\begin{footnotesize}
%%\begin{spacing}{1.0}
%The work is supported by the
%National Science Foundation of China under Grant No. $\cdots$
%%\end{spacing}
%%\end{footnotesize}
%%The authors would like to thank...

% trigger a \newpage just before the given reference
% number - used to balance the columns on the last page
% adjust value as needed - may need to be readjusted if
% the document is modified later
%\IEEEtriggeratref{8}
% The "triggered" command can be changed if desired:
%\IEEEtriggercmd{\enlargethispage{-5in}}

% references section

% can use a bibliography generated by BibTeX as a .bbl file
% BibTeX documentation can be easily obtained at:
% http://www.ctan.org/tex-archive/biblio/bibtex/contrib/doc/
% The IEEEtran BibTeX style support page is at:
% http://www.michaelshell.org/tex/ieeetran/bibtex/
\bibliographystyle{IEEEtran}
% argument is your BibTeX string definitions and bibliography database(s)
%\bibliography{IEEEabrv,../bib/paper}
%
% <OR> manually copy in the resultant .bbl file
% set second argument of \begin to the number of references
% (used to reserve space for the reference number labels box)

%\bibliographystyle{plain}
%\begin{thebibliography}{99}
%
%
%\end{thebibliography}
%

\bibliography{bibfile}

% that's all folks
\end{document}